\documentclass[pre,twocolumn]{revtex4-1}

\usepackage{graphicx}
\usepackage{amsmath}
\usepackage{mathtools}
\usepackage{amssymb}

\usepackage{mathtools,cancel}
\usepackage{graphicx} 
\usepackage{booktabs} 
\usepackage{MnSymbol}

\usepackage{verbatim}   
\usepackage{color}      
\usepackage{subfigure}  
\usepackage{hyperref}   

\newcommand{\ba}{\begin{eqnarray}}
\newcommand{\ea}{\end{eqnarray}}
\newcommand{\be}{\begin{equation}}
\newcommand{\ee}{\end{equation}}

\begin{document}

\title{Two-component Gaussian core model:  strong-coupling limit, Bjerrum pairs, and gas-liquid phase transition}

\author{Derek Frydel}
\affiliation{Federico Santa Mar\'ia Technical University, 
San Joaqu\'in, Santiago, Chile \\
Institute of Physics, The Federal University of Rio Grande do Sul, Porto Alegre 91501-970, Brazil }
\email{dfrydel@gmail.com}
\author{Yan Levin}
\affiliation{Institute of Physics, The Federal University of Rio Grande do Sul, Porto Alegre 91501-970, Brazil}

\date{\today}

\begin{abstract}
In the present work we investigate a gas-liquid transition in a two-component Gaussian core model, 
where particles of the same species repel and those of different species attract.  Unlike a similar transition 
in a one-component system with particles having attractive interactions at 
long separations, and repulsive interactions at short separations, a transition in the two-component 
system is not driven solely by interactions, but by a specific feature of the interactions, the correlations. 
This leads to extremely low critical temperature, as correlations are dominant in the strong-coupling limit.  
By carrying out various approximations based on standard liquid-state methods, we show that a gas-liquid 
transition of the two-component system posses a challenging theoretical problem.  
\end{abstract}

\pacs{
}

\maketitle

\section{Introduction}

A two-component fluid with interactions 
\be
u_{ij}(r) = \left\{ 
  \begin{array}{r l}
     u(r), & \quad \text{if $i=j$}\\
    -u(r), & \quad \text{if $i\ne j$,}
        \label{eq:u0}
  \end{array} 
  \right.
  \ee
where the indices designate the two components, is best exemplified by electrolytes.  
Both in experiments and idealized representations, such as primitive models \cite{Yan93,Yan96,Evans96,Pan05,Pan08} 
or penetrable ions  \cite{Hansen11a,Hansen11b,Hansen12,Frydel12,Masters13,Levesque14,Frydel16a,Frydel16b}, 
electrolytes undergo a gas-liquid transition.   
A similar transition is expected for any two-component fluid as long as interactions obey Eq. (\ref{eq:u0}), 
and it can be understood by adopting an effective one-component description, where 
particles of one species experience mediated attractive 
interactions due to averaged contributions of the second species.  
The attractions eventually lead to phase transition.  

A theoretical challenge posed by the models conforming to Eq. (\ref{eq:u0}) is that they 
can only be described by a theoretical framework containing correlations 
\cite{Yan02}, therefore, a straightforward application of mean-field techniques is of no use.  
The simplest theory of correlations is the random phase approximation (RPA), which is equivalent
to a one-loop expansion around the mean-field solution (or the saddle-point) \cite{Frydel16,Frydel17}.  
Corrections due to the strong-coupling limit effects
can be incorporated into the RPA framework by an explicit incorporation of 
Bjerrum pairs, which are dimers between particles of opposite species 
\cite{Ebeling68,Ebeling71,Yan93,Yan96,Hansen12}.  
For the primitive model this procedure 
correctly shifts the critical point of a gas-liquid transition 
to higher densities.  On the other hand, for penetrable ions the same procedure leads to no 
satisfactory results \cite{Hansen12}.



Motivated by this theoretical difficulty of treating fluids with the binary interactions of the form presented 
in Eq. (\ref{eq:u0}), the present article considers a fluid of Gaussian particles, the so-called Gaussian core 
model (GCM).  Like penetrable ions, Gaussian particles are penetrable, but unlike penetrable ions, 
Gaussian interactions are short-ranged.  
A one-component GCM fluid has been studied extensively in the past by numerous groups 
\cite{Stillinger76,Stillinger78a,Stillinger78b,Stillinger79a,Stillinger79b,Stillinger97,Hansen00}.  
The two-component version of the GCM model, but for repulsive-only interactions, was investigated 
in \cite{Evans01}.  The two-component GCM fluid with interactions satisfying Eq. (\ref{eq:u0}) was 
briefly introduced in \cite{Frydel16} as a testing ground for the generalized-RPA approximation.





This work is organized as follows.  In Sec. (\ref{sec:GCM}) we introduce the GCM model.  In 
Sec. (\ref{sec:simulation}) we present the simulation results, focusing on the location of the critical 
point and the structure of a fluid with special view to dimer formation.  In Sec. (\ref{sec:theory}) we 
analyze the GCM model using a number of approximations.  
Finally, in Sec. (\ref{sec:concl}) we conclude the work.

\section{The Gaussian core model}
\label{sec:GCM}
Particles in the GCM model interact via the Gaussian potential, $u(r)=\epsilon e^{-r^2/\sigma^2}$, 
and for the two-component system considered in the present work the interactions are 
\be
u_{ij}(r) = \left\{ 
  \begin{array}{r l}
     \varepsilon e^{-r^2/\sigma^2}, & \quad \text{if $i=j$}\\
    -\varepsilon e^{-r^2/\sigma^2}, & \quad \text{if $i\ne j$,}
        \label{eq:u}
  \end{array} 
  \right.
  \ee
where the indices $i,j=1,2$ designate different species, $\varepsilon$ is the depth of the potential, 
and $\sigma$ is the particle diameter.  In the following, we use the physical quantities reduced by 
$\sigma$, $\varepsilon$, and the Boltzmann constant $k_B$.  The reduced length is $r^*=r/\sigma$, 
and the reduced density is $\rho^*=\rho\sigma^3$.  Then the reduced temperature is 
$T^*=k_BT/\varepsilon$, the reduced pressure is $P^*=P\sigma^3/(k_BT)$, and the strength of the 
Gaussian potential in relation to thermal energy is $\varepsilon^*= 1/T^* = \varepsilon/(k_BT)$.

Simulation results in the present work are from the standard canonical Monte Carlo simulations, 
$(N,V,T)$.
Because the pair potentials of the GCM fluid are bounded, there is no constraint for the 
displacement length of attempted moves, which could be very long.  In our simulations we set
this length to ensure that the acceptance ratio is larger than $50\%$ and less than $80\%$.
All simulations are 
performed in a cubic box with periodic boundary conditions in all three directions.  
The total number of particles in the box is $N_1+N_2=1000$, and the bulk densities $\rho_1$
and $\rho_2$ are controlled by the box size $L$.  As $N_1=N_2=N/2$, the density of 
each species is $\rho_1=\rho_2=\rho/2$.

\section{Simulation results}
 \label{sec:simulation}

\subsection{Pressure Isotherms and the Critical Point}
\begin{figure}[h] 
 \begin{center}
 \begin{tabular}{rr}
  \includegraphics[height=0.21\textwidth,width=0.27\textwidth]{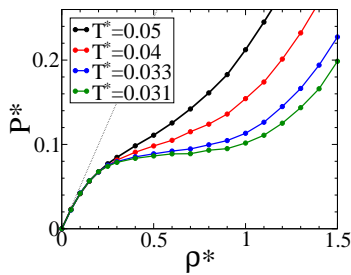}
 \end{tabular}
 \end{center}
\caption{Pressure isotherms for the two-component GCM fluid generated by  MC simulations.  
The dashed line designates an ideal-gas pressure for 
$N/2$ dimers, $P^*=\rho^*/2$.  } 
\label{fig:P_sim}
\end{figure}

The presence and location of the critical point of a gas-liquid phase transition is determined 
from pressure isotherms, where we know that the critical temperature isotherm includes a stationary 
inflection point, $\left(\frac{\partial P}{\partial\rho}\right)_T = \left(\frac{\partial^2 P}{\partial\rho^2}\right)_T =0$. 
Fig. (\ref{fig:P_sim}) displays a number of isotherms generated by simulations.  
A gradual emergence of a plateau in the shape of an isotherm with decreasing temperature 
indicates the approaching critical temperature.  The critical point is estimated to be 
roughly at $T^*_c\approx 0.03$ and $\rho_c^*\approx 0.6$.    
\graphicspath{{figures/}}

Such a low critical temperature is a signature of a phase transition that is driven 
by correlations.  Other systems with the binary interactions of the form presented in Eq. (\ref{eq:u0})
exhibit a similar behavior.  For example, the critical temperature of penetrable ions is 
$T_c^*\approx 0.02$ \cite{Hansen11b}, and that of restrictive primitive model is 
$T_c^*\approx 0.07$ \cite{Yan93}.  In contrast, the critical temperature of an analogous 
gas-liquid phase transition of a one-component Lennard-Jones fluid is $T_c^*\approx 1.1$ 
\cite{Hu12}, which roughly corresponds to an equivalence between the thermal and the potential 
energy at minimum. 

We note that normally a critical point depends on a size of a simulation box $L$ \cite{Levesque14}.  
For the critical temperature this dependence is expressed as $T_c(L) = T_c + L^{-a}$, where $T_c$ is the
critical temperature in the thermodynamic limit, and the value of $a$ for the Ising model is $\sim 2.44$.  
In the present work 
we are interested in general properties of a fluid prior to the onset of a phase transition, and a rough
estimate of the critical point is sufficient for our purposes.  To check finite size effects, 
we generate a number of isotherms for $N=2000$ particles, but we find no change
in the results.

\subsection{Correlation Functions}
A unique feature of a two-component system with the binary interactions of the form presented in 
Eq. (\ref{eq:u0}) is the formation of dimers between particles of opposite species, known as Bjerrum 
pairs in the context of electrolytes.  At low temperature and density these pairs dominate the fluid 
structure and, because of their stability, can be regarded as a third component.  It is not clear, however, 
what effect, if any, the pairs may have on a phase transition, and wether a successful theory of a 
phase-transition is required to incorporate pair formation.  

\begin{figure}[h] 
 \begin{center}
 \begin{tabular}{rrrrrr}
  \includegraphics[height=0.17\textwidth,width=0.22\textwidth]{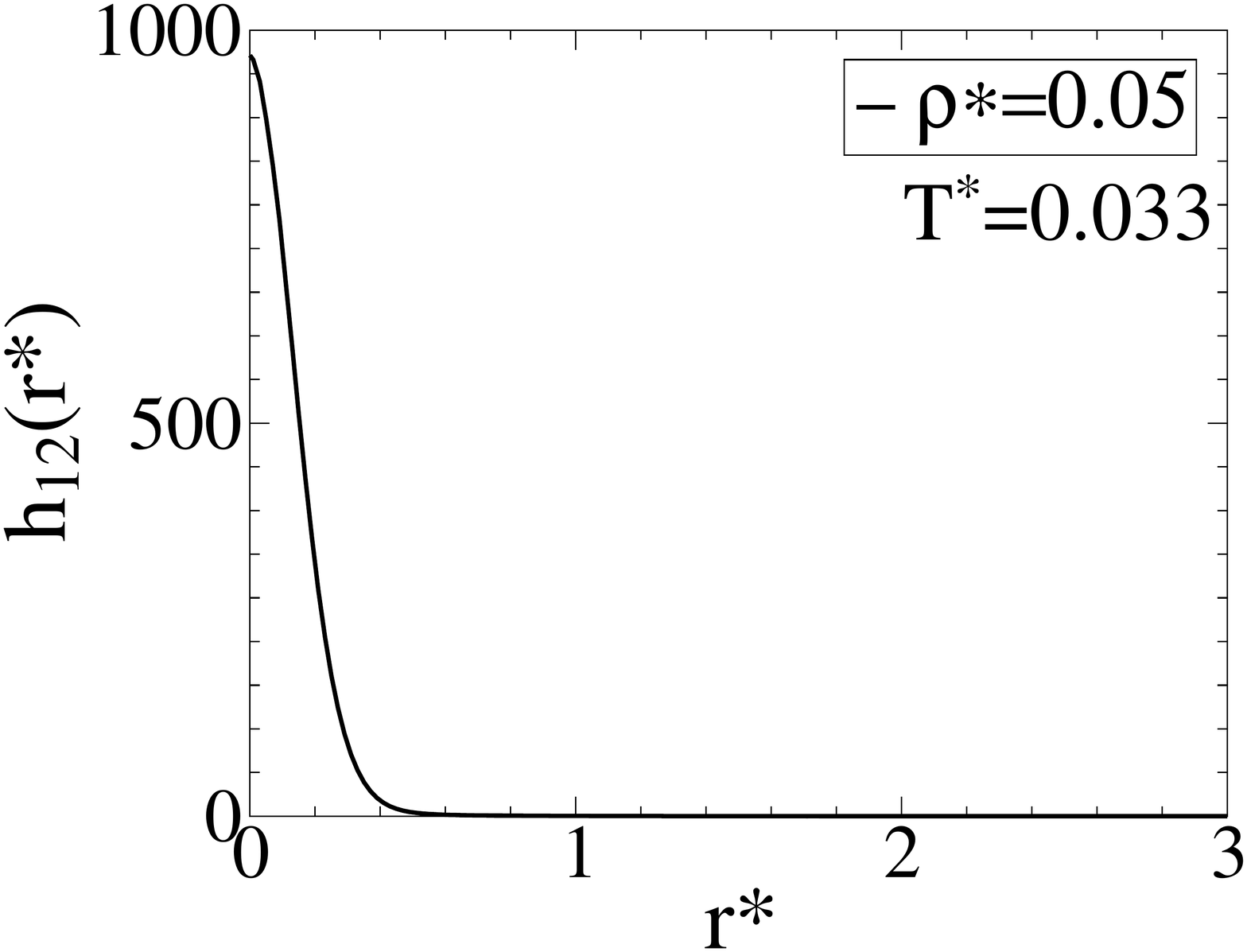}&
  \includegraphics[height=0.17\textwidth,width=0.22\textwidth]{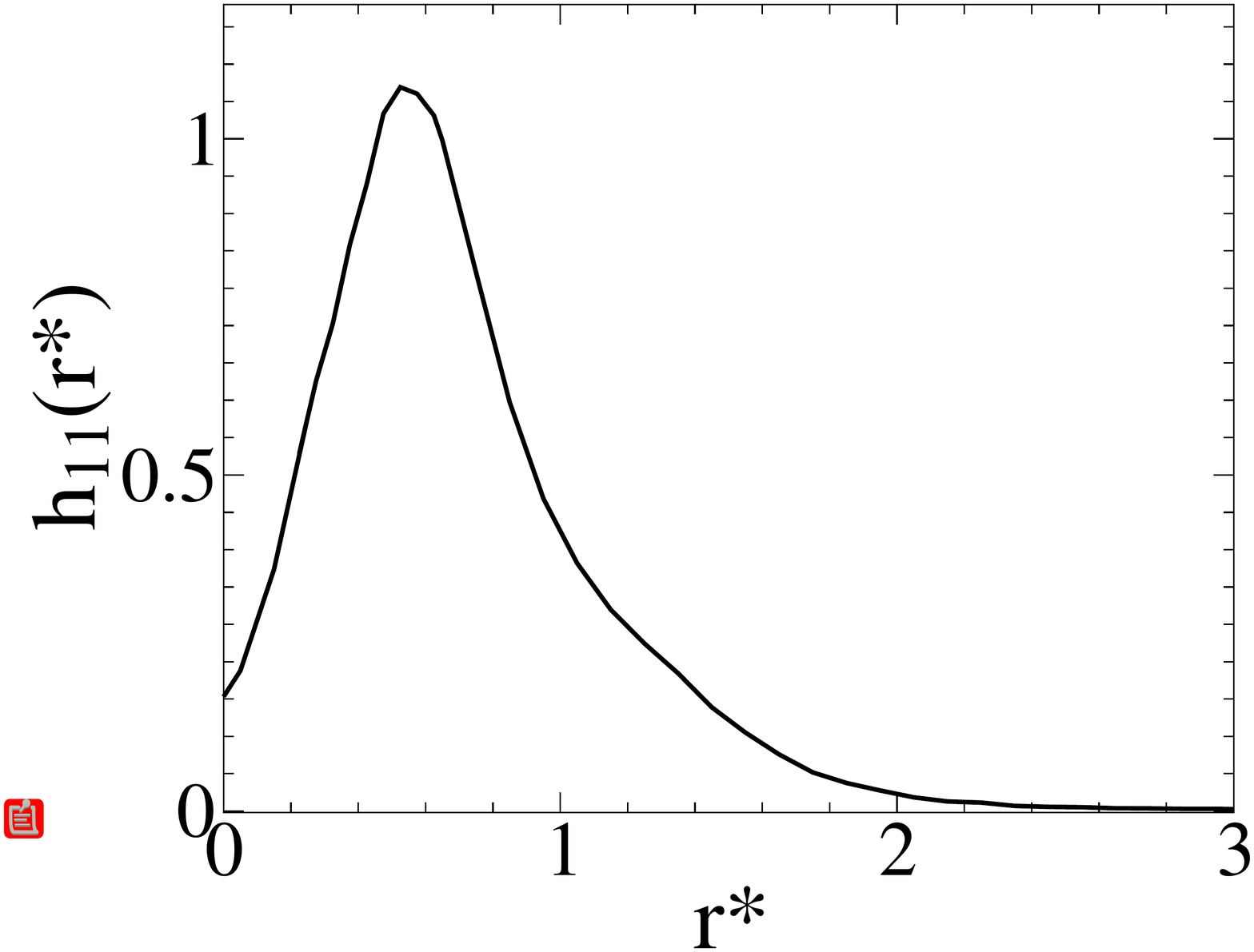}\\
  \includegraphics[height=0.17\textwidth,width=0.22\textwidth]{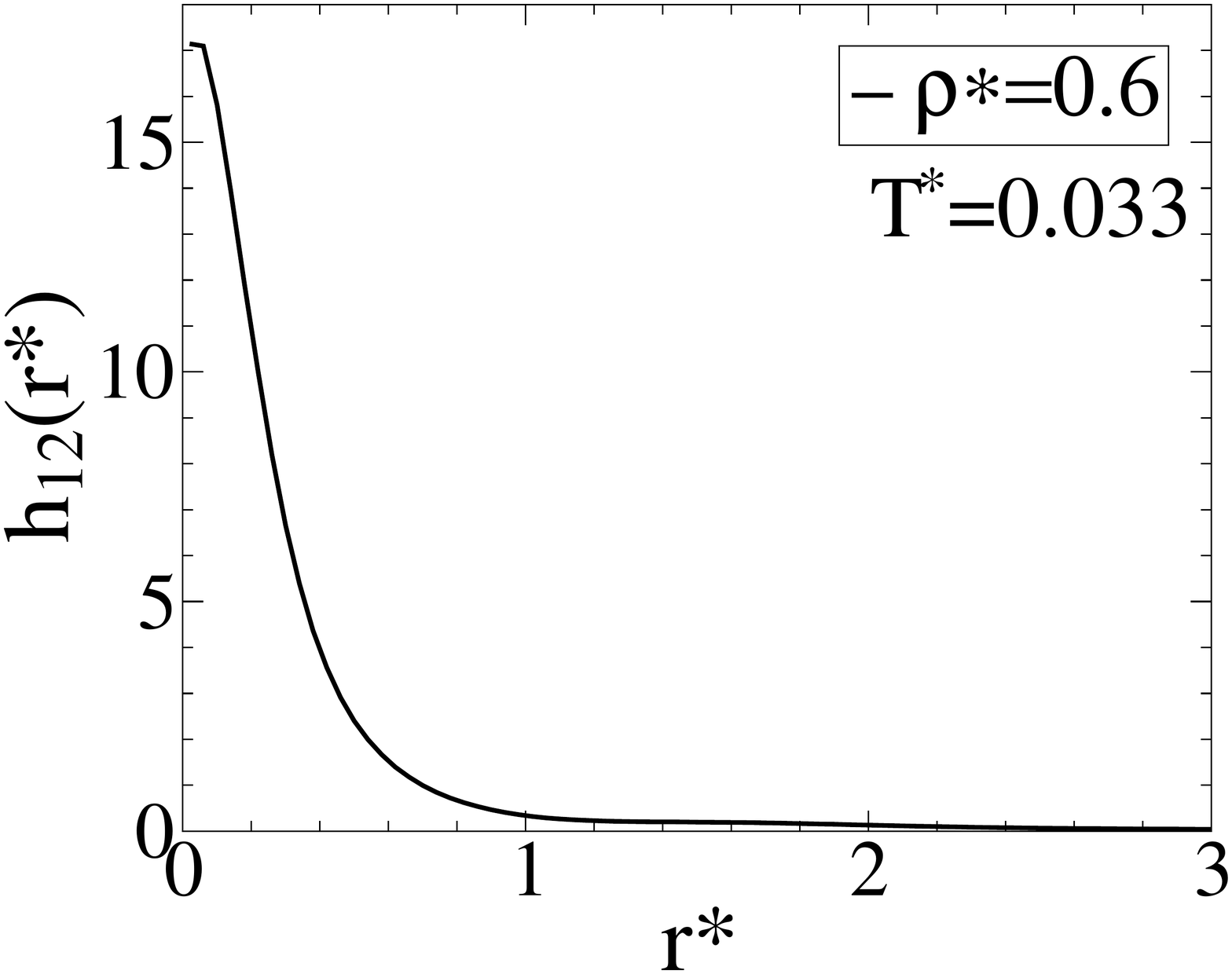}&
  \includegraphics[height=0.17\textwidth,width=0.22\textwidth]{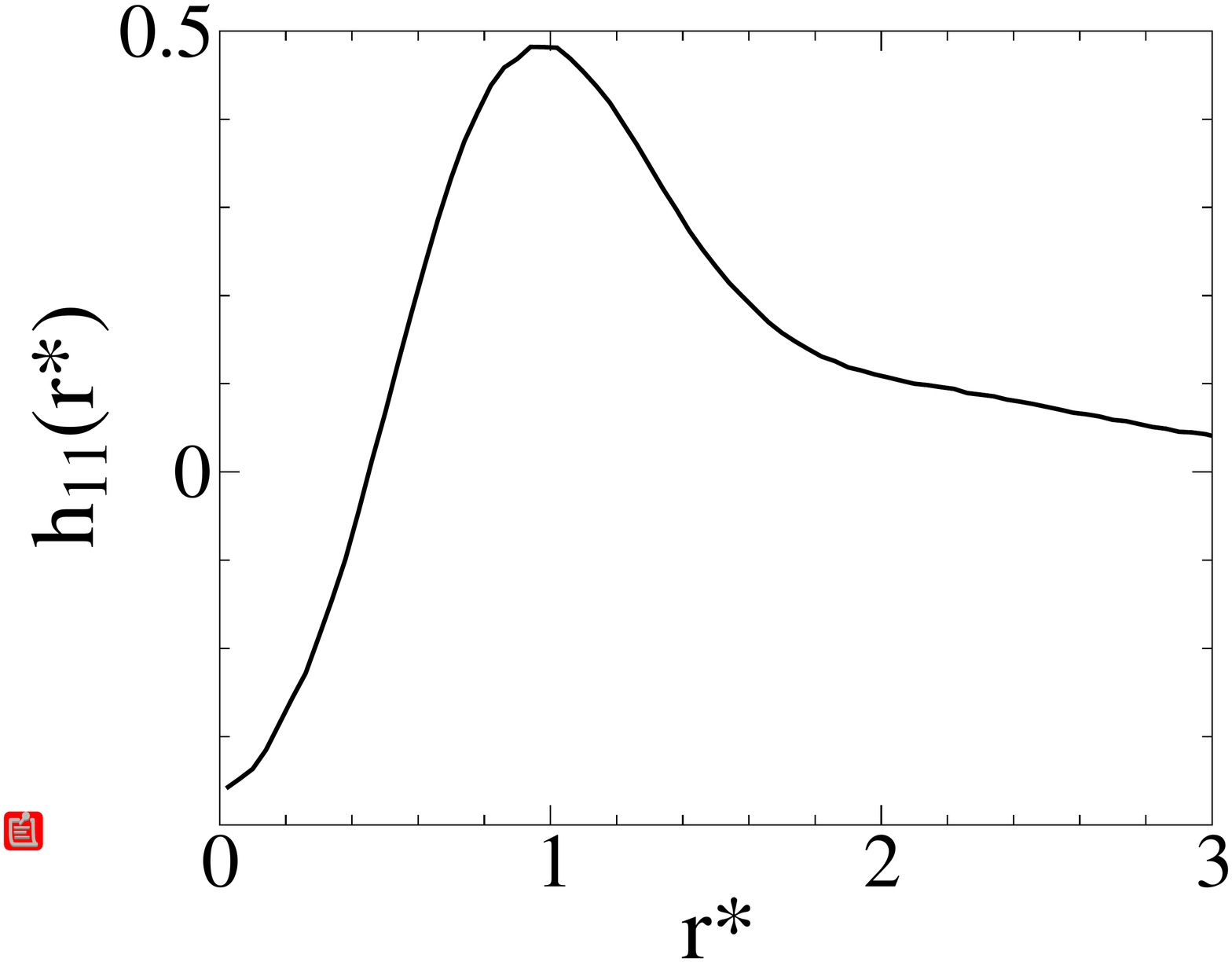}\\
  \includegraphics[height=0.17\textwidth,width=0.22\textwidth]{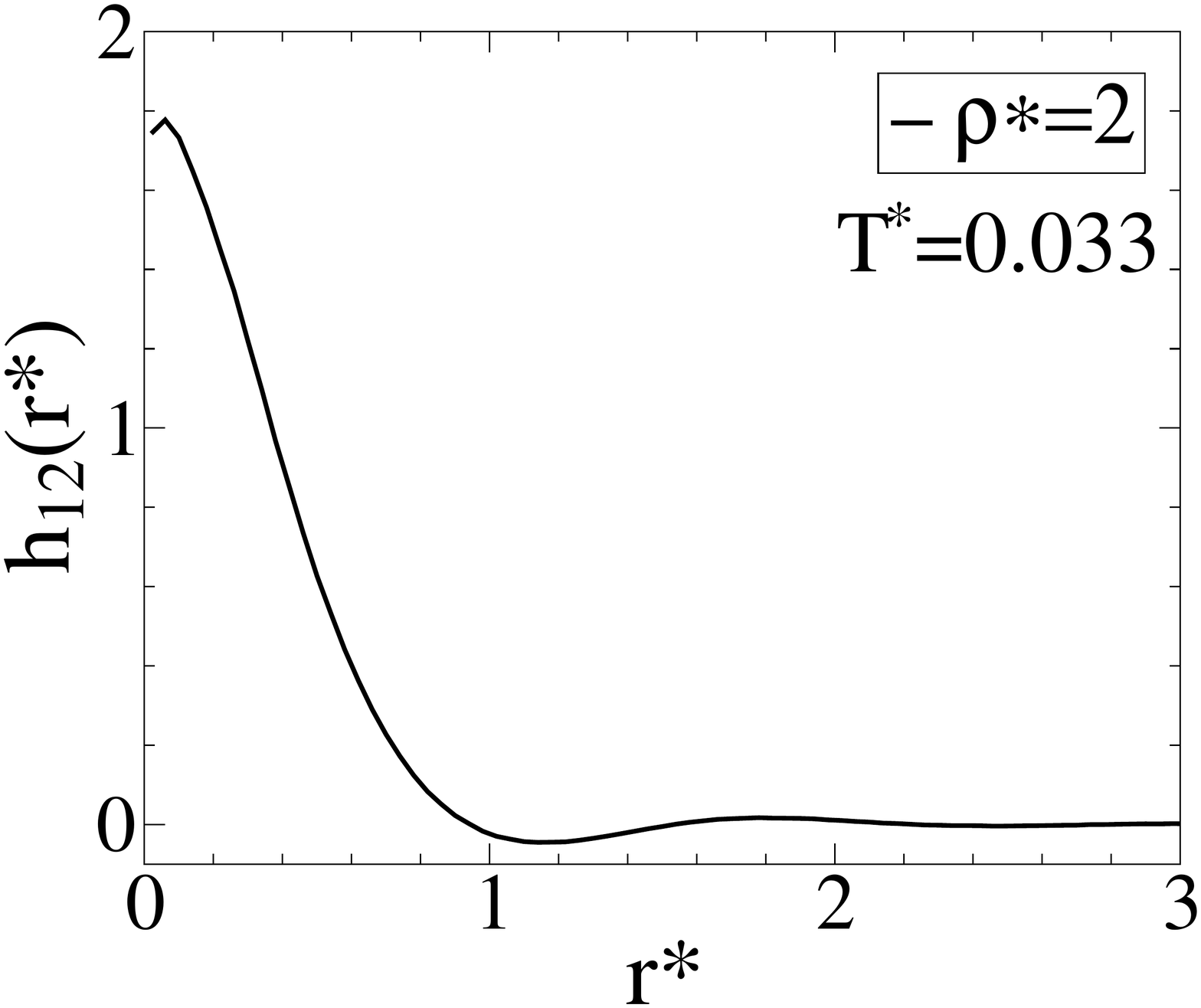}&
  \includegraphics[height=0.17\textwidth,width=0.22\textwidth]{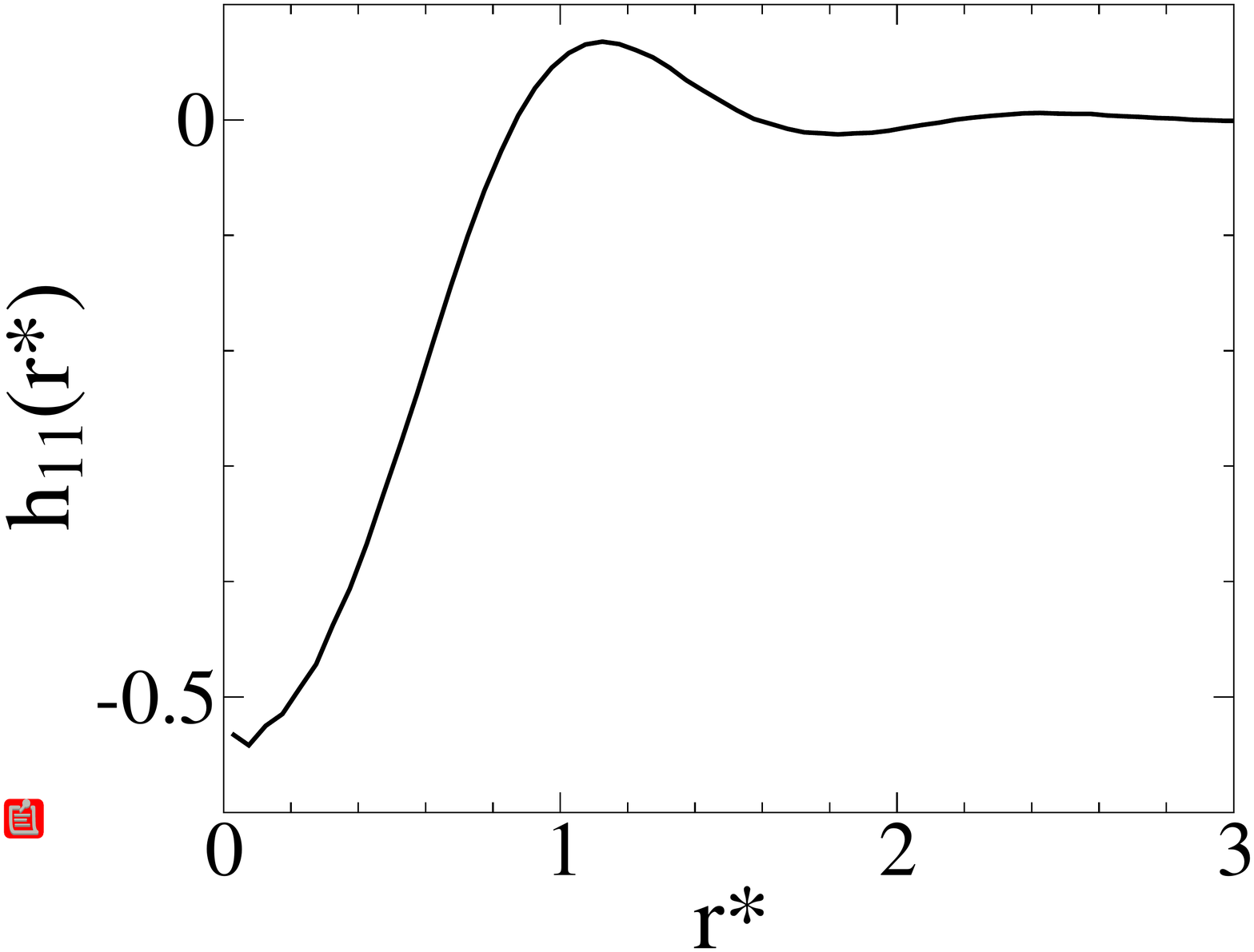}\\
 \end{tabular}
 \end{center}
\caption{Pair correlation functions, $h_{11}(r)$ and $h_{12}(r)$, for the two-component GCM fluid for 
$T^*=0.033\gtrsim T^*_c$ for three densities $\rho^*=0.05, 0.6, 2$. } 
\label{fig:h_sim}
\end{figure}

To examine this question, in this section we consider a number of correlation functions.  
The two relevant correlations are:  correlations between particles of the same species, $h_{11}(r)$, 
and correlations between particles of different species, $h_{12}(r)$.  In Fig. (\ref{fig:h_sim}) we plot 
these functions for different densities slightly above the critical temperature, $T^*\gtrsim T_c^*\approx 0.03$.    
\graphicspath{{figures/}}

First, we consider the function $h_{12}(r)$.    
For the lowest density, the correlation function is dominated  
by a sharp peak at $r=0$ and the absence of oscillations, indicating an absence of the 
secondary structures.  This suggests that particles exist as pairs, and the system can be regarded 
as an ideal-gas of $N/2$ dimers.  This is confirmed by an incipient 
agreement  in Fig. (\ref{fig:P_sim}) between the pressure isotherms generated by simulations 
and the linear behavior $P^*=\rho^*/2$.  Dimers disintegrate with increasing density, which is 
seen as $h_{12}(r)$ develops a usual oscillatory structure.

We examine next the function $h_{11}(r)$.  
Because the bare interactions between particles of the same species are repulsive, $h_{11}(r)$ is 
expected to feature a correlation hole, a region of negative correlations around $r=0$.  This 
expectation is satisfied for densities $\rho^*=0.6$ and $\rho^*=2$.  But for $\rho^*=0.05$ the 
correlation hole vanishes and is substituted by a region of positive correlations, indicating effective 
attractions.  Since from $h_{12}(r)$ we know that dimers dominate a fluid structure at $\rho^*=0.05$, 
the effective attractions imply attractions between dimers.  This, in turn, explains a gradual deviation 
of pressure isotherms from the ideal-gas 
behavior $P^*=\rho^*/2$, suggesting a corrected low density ansatz $P^*=\rho^*/2 + B_2\rho^{*2}$.

\begin{figure}[h] 
 \begin{center}
 \begin{tabular}{rr}
  \includegraphics[height=0.21\textwidth,width=0.27\textwidth]{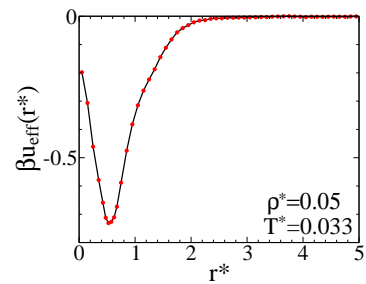}
 \end{tabular}
 \end{center}
\caption{Mediated interactions between particles of the same species defined as $\beta u_{\rm eff}(r)=-\log h_{11}(r)$, for
density $\rho^*=0.05$ and temperature $T^*=0.033\gtrsim T^*_c$. } 
\label{fig:u11_sim}
\end{figure}

The above discussion suggests that it may be interesting to adopt 
an effective one-component description by integrating out the degrees of freedom
of the second component, analogous to the concept of depletion interactions \cite{Roland99,Roland00,Goulding01}.  
At low density effective interactions can be obtained by inverting a pair correlation function,  
$\beta u_{\rm eff}=-\log h_{11}(r)$, shown in Fig. (\ref{fig:u11_sim}) for $\rho^*=0.05$ and $T^*=0.033$.  
The resulting effective potential is everywhere negative with a minimum at $r^*\approx 0.5$.  
We attempt next to carry out a simulation for a one-component system for $\rho^*=0.025$ and interactions 
$u_{\rm eff}(r)$.  We find this system thermodynamically unstable, with all particles collapsing into a single cluster 
with every particle being in overlap with all others.  
To make the description more realistic and prevent such a collapse, 
one would need to include three-body and perhaps higher-body effective 
interactions.  This, however, greatly complicates the required computations and, in effect, makes 
the effective one-component approach unfeasible.  
\graphicspath{{figures/}}

\subsection{Pairs}
Another way of looking at a fluid structure, and, especially, formation of dimers, is by analyzing 
the quantity 
\be
C_{ij}=\rho_i \int_0^{\infty}dr\, 4\pi r^2 h_{ij}(r), 
\label{eq:Cij}
\ee
where $\rho_i h_{ij}(r)$ is a density perturbation of a species ``$i$" around 
a fixed fluid particle of a species ``$j$".  (We recall that $\rho_i=\rho/2$).  

\begin{figure}[h] 
 \begin{center}
 \begin{tabular}{rr}
 \hspace{-0.5cm}
  \includegraphics[height=0.19\textwidth,width=0.24\textwidth]{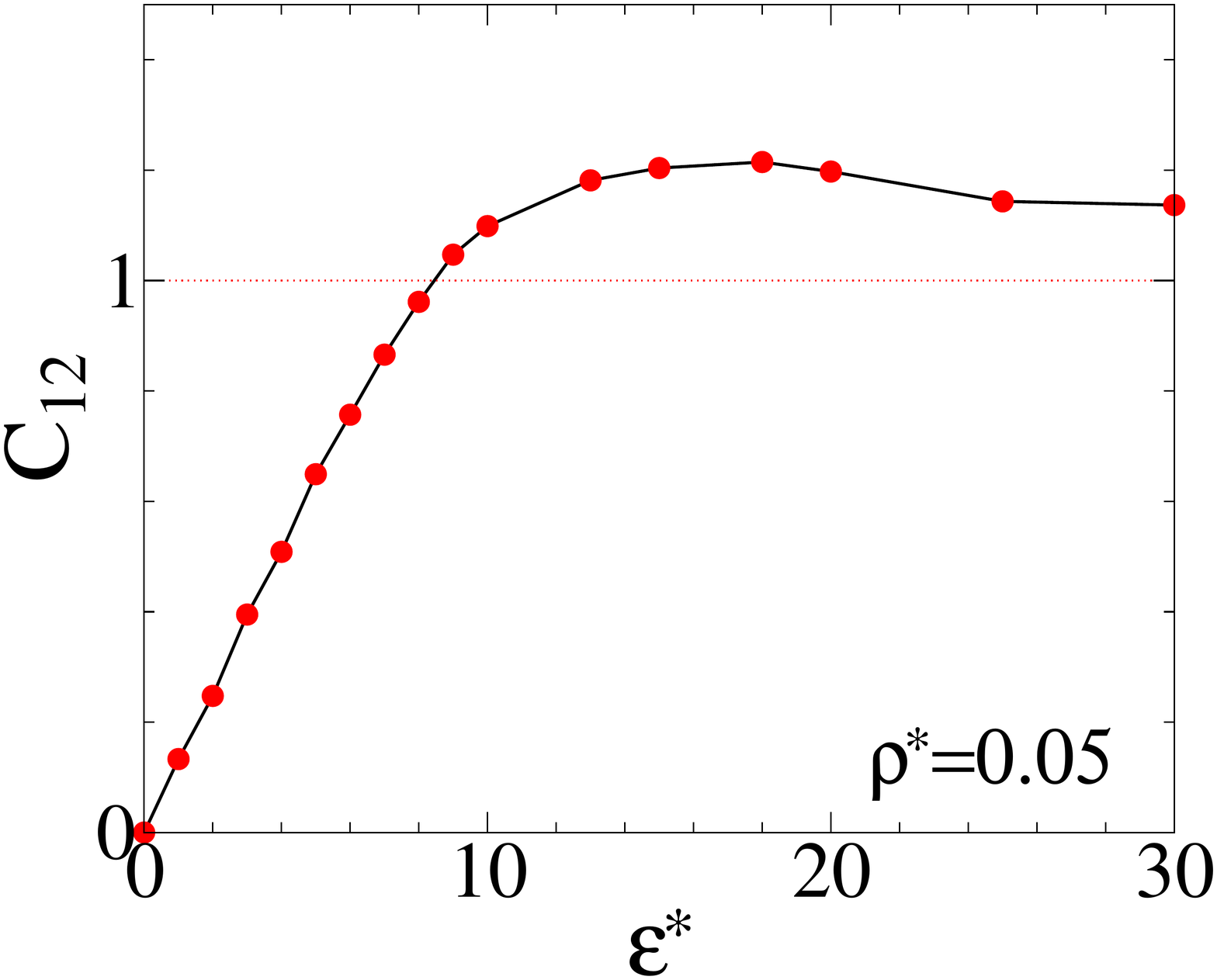}&
  \includegraphics[height=0.19\textwidth,width=0.24\textwidth]{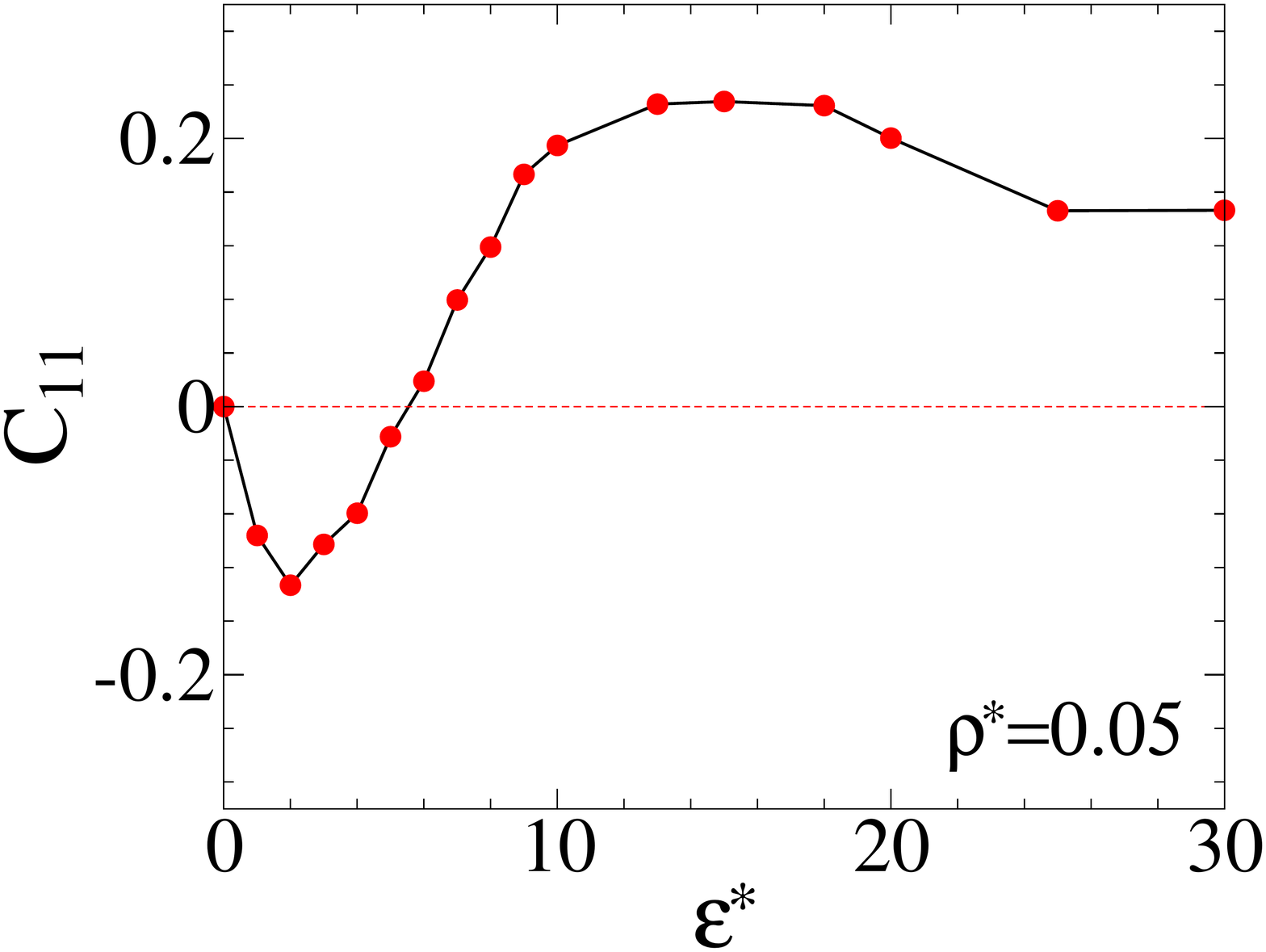}
 \end{tabular}
 \end{center}
\caption{The quantity $C_{ij}=\rho_i \int_0^{\infty}dr\, 4\pi r^2 h_{ij}(r)$ as a function 
of $\varepsilon^*=1/T^*$ for $\rho^*=0.05$, prior to the onset of phase transition.  
The data points are from simulations.  
} 
\label{fig:C1}
\end{figure}

The data points obtained from simulations for a low density $\rho^*=0.05$ are plotted in Fig. (\ref{fig:C1}) as 
a function of $\varepsilon^*=1/T^*$, prior to the onset of a phase transition.  To better understand the 
quantity $C_{12}$ in Fig. (\ref{fig:C1}) (a), 
it is helpful to keep in mind the shape of a correlation function $h_{12}(r)$ for the same density in 
Fig. (\ref{fig:h_sim}) (a).  The emergence of quasi-stable pairs corresponds to $C_{12}\approx 1$.  
Initially $C_{12}$ increases linearly, and at $\varepsilon^*=10$ (or $T^*=0.1$) all particles are paired. 
However, instead of saturating at $1$, $C_{12}$ continues to increase as a consequence of mediated 
attractive interactions, indicating that pairs are not ideal-gas particles.  
\graphicspath{{figures/}}

Next, we interpret the data points for $C_{11}$ in Fig. (\ref{fig:C1}) (b).  In case of weak interactions 
$C_{11}$ is negative, reflecting the presence of a correlation hole in $h_{11}(r)$.  These negative 
correlations begin do disappear around $\varepsilon^*\approx 2$, then at $\varepsilon^*\approx 6$
the quantity $C_{11}$ changes sign and becomes positive, indicating the onset of effective attractive 
interactions.  

In analogy to Coulomb particles, we consider next 
the quantity 
\be
C_d=\rho_i \int_0^{\infty}dr\, 4\pi r^2 \left[h_{ii}(r)-h_{ij}(r)\right] \ge -1, 
\label{eq:hc_hole}
\ee
where $C_d$ designates a "charge" that a fixed fluid particle attracts.  
In contrast, for a Coulomb system $C_d=-1$, as a consequence of long-range interactions, and implies
the perfect screening.  This exact 
condition is referred to as the (zero-order) Stillinger-Lovett sum rule \cite{Lovett68a,Lovett68b,Martin88}.  
On the other hand, the GCM two-component fluid achieves perfect screening gradually as $T^*\to 0$
or $\rho^*\to\infty$.  
\graphicspath{{figures/}}
\begin{figure}[h] 
 \begin{center}
 \begin{tabular}{rr}
 \hspace{-0.3cm}
  \includegraphics[height=0.19\textwidth,width=0.24\textwidth]{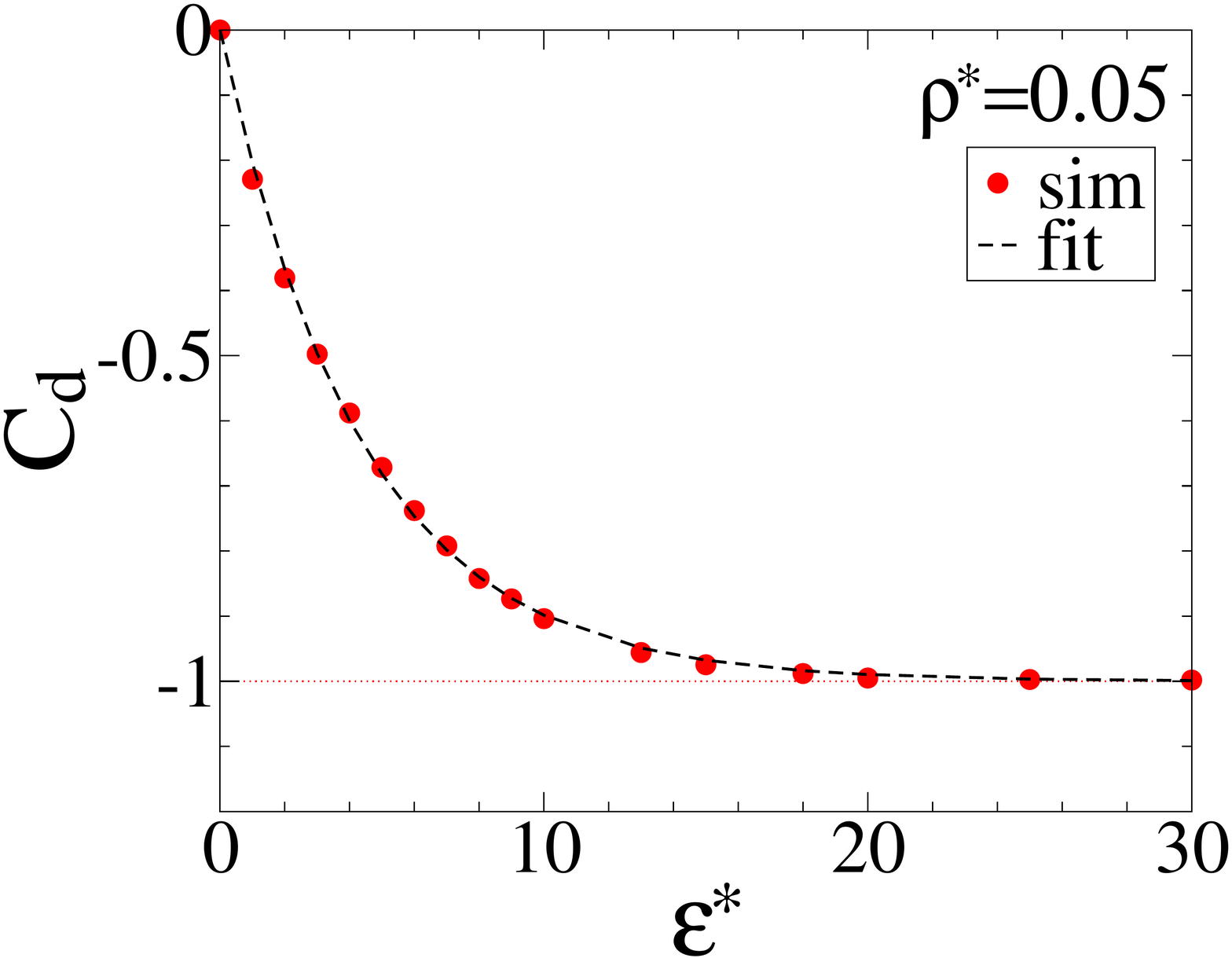}&
  \hspace{-0.3cm}
  \includegraphics[height=0.19\textwidth,width=0.24\textwidth]{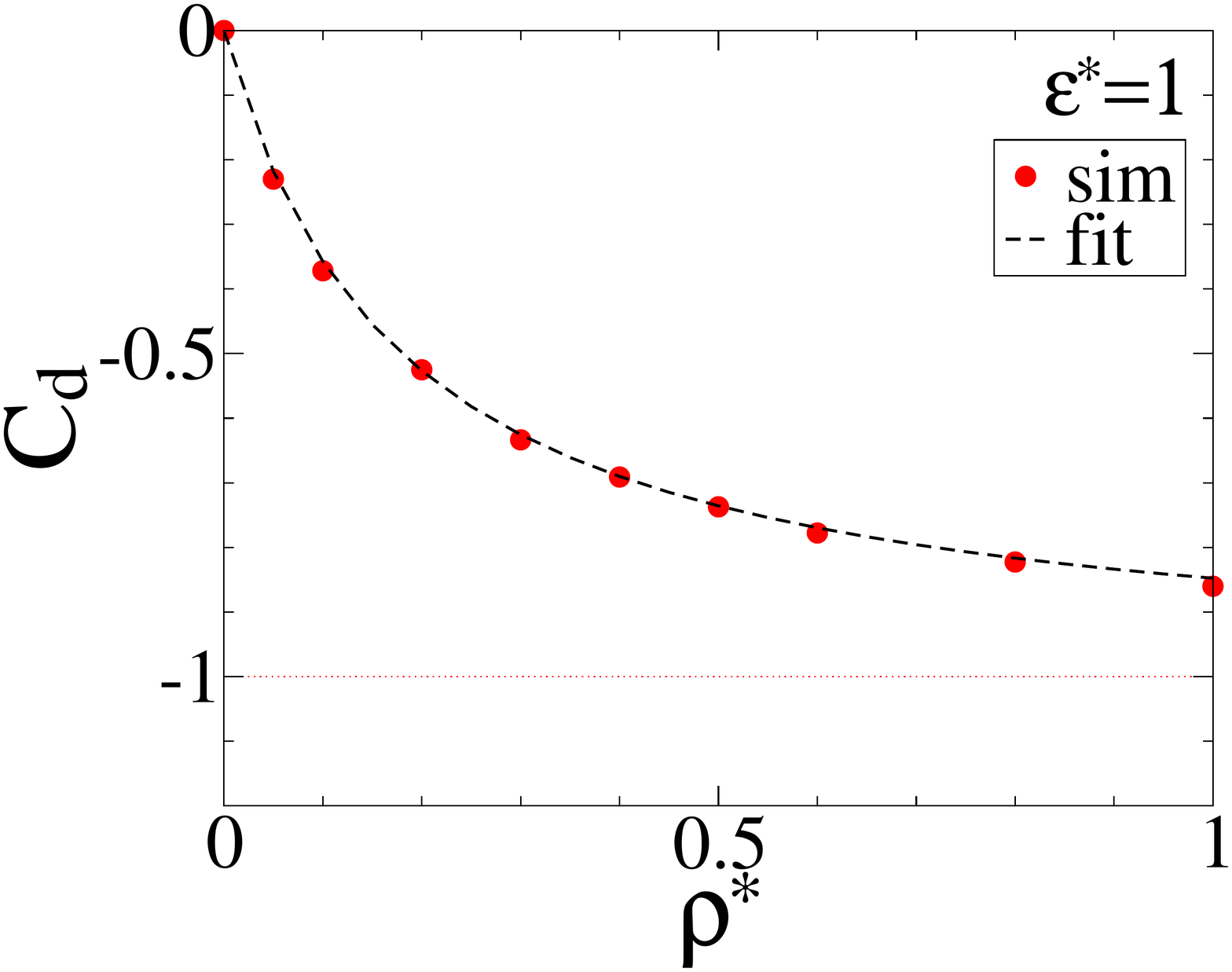}
 \end{tabular}
 \end{center}
\caption{The quantity $C_d= \rho_i \int_0^{\infty}dr\, 4\pi r^2 \big\{h_{ii}(r)-h_{ij}(r)\big\}$, designating 
the total ``charge" a fixed particle attracts, (a) as a function of $\varepsilon^*$ for $\rho^*=0.05$, 
and (b) as a function of $\rho^*$ for $\varepsilon^*=1$.  $C_d$ is bounded from below by $-1$. 
The fits to the data points for respective plots are $C_d=e^{-\varepsilon^*/\varepsilon_0^*}-1$, with $\varepsilon_0^*\approx 4.3$, 
and $C_d=\frac{\pi^{3/2}\varepsilon^*\rho^*}{1+\pi^{3/2}\varepsilon^*\rho^*}$ (obtained from the RPA approximation).} 
\label{fig:C}
\end{figure}

The plots of Fig. (\ref{fig:C}) are fitted to simple functional forms.  
A fast exponential decay in Fig. (\ref{fig:C}) (a) 
agrees with the notion that dimers dominate the fluid structure at low densities for $T^*< 0.1$.   
$C_d$ as a function of $\rho^*$ and for $T^*=1$ in Fig. (\ref{fig:C}) (b), on the other hand,  
can be fit to an algebraic decay obtained from the RPA approximation, which is a weak-coupling theory, 
and does not capture the formation of dimers.

\section{Theoretical analysis}
\label{sec:theory}
\subsection{Random Phase Approximation}
\label{sec:rpa}

The free energy density for the two-component system with interactions in Eq. (\ref{eq:u0})
has two main contributions,  
\be
f = f_{\rm id} + f_c,
\ee
where the ideal-gas contribution is 
\be
f_{\rm id} = k_BT\rho \left(\log\frac{\rho\Lambda^3}{2} - 1\right),
\ee
($\Lambda$ is a de Broglie wavelength)  
and the expression for the correlation free energy density 
can be obtained from the adiabatic connection \cite{Frydel16,Frydel17}, 
wherein the pair interaction $\lambda u(r)$ is gradually turned on by changing $\lambda$ from 
$0\to 1$.  The resulting expression for the present two-component homogeneous system is 
\be
f_c = \pi\rho^2 \int_0^{\infty} dr \, r^2 u(r) \int_0^1 d\lambda\,h_d^{\lambda}(r),
\label{eq:hd}
\ee
where $h_d^{\lambda}(r)=h_{11}^{\lambda}(r)-h_{12}^{\lambda}(r)$ and the superscript $\lambda$
indicates the correlation function for particles with interactions $\lambda u(r)$.  


The correlation function in Eq. (\ref{eq:hd}) is obtained from the Ornstein-Zernike equation (OZ), 
\be
h_{ij}^{\lambda}(r) = -\beta c_{ij}^{\lambda}(r) - \beta\sum_{k=1}^2\rho_k\int d{\bf r}'\,h_{kj}^{\lambda}(r') 
c_{ik}^{\lambda} (|{\bf r}'-{\bf r}|), 
\label{eq:OZ}
\ee
where within the RPA approximation the direct correlation function is approximated as 
$c_{ij}^{\lambda,{\rm rpa}}(r)=-\lambda\beta u_{ij}(r)$.  This in turn implies that  
$h^{\lambda,{\rm rpa}}_{11}(r)=-h^{\lambda,{\rm rpa}}_{12}(r)=h^{\rm rpa}_{\lambda}(r)$, where the function 
$h^{\rm rpa}_{\lambda}(r)$ is obtained from the modified OZ equation, 
\be
h_{\lambda}^{\rm rpa}(r) + \beta\lambda u(r) = - \beta\lambda \rho\int d{\bf r}'\,h_{\lambda}^{\rm rpa}(r')u(|{\bf r}'-{\bf r}|), 
\label{eq:h_rpa}
\ee
leading to the approximate correlation free energy, 
\be
\beta f_{\rm c}^{\rm rpa} = -\frac{\rho}{2}\int_0^1 d\lambda\, \frac{h_{\lambda}^{\rm rpa}(0)+\lambda \beta u(0)}{\lambda}.  
\ee

To obtain analytical results, Eq. (\ref{eq:h_rpa}) is Fourier transformed 
\be
\hat h_{\lambda}^{\rm rpa}(k) = -\frac{\beta\lambda \hat u(k)}{1+\beta\lambda\rho \hat u(k)}. 
\ee
After the Fourier inversion, the correlation free energy becomes 
\be
\beta f_{\rm c}^{\rm rpa} = \frac{1}{4\pi^2}\int_0^{\infty} dk\, k^2 \bigg(\log\big[1+\rho \beta \hat u(k)\big] - \rho \beta \hat u(k)\bigg), 
\label{eq:fc1}
\ee
which after substitution $\hat u(k)=\sigma^3\pi^{3/2}e^{-k^2\sigma^2/4}$ evaluates to   
\be
\beta f_{\rm c}^{\rm rpa}  
= -\frac{\varepsilon^*\rho}{2}\bigg\{1+\frac{{\rm Li}_{5/2}[-\varepsilon^*\eta]}{\varepsilon^*\eta}\bigg\},
\ee
where $\eta = (\sigma\sqrt{\pi})^3\rho$, and ${\rm Li}_m(x) = \sum_{n=1}^{\infty}\frac{x^n}{n^{m}}$ 
is a polylogarithm.

To locate the critical point, we study pressure isotherms.  The expression of pressure
is obtained from either the thermodynamic definition 
$P=-\frac{\partial F}{\partial V} = \rho\frac{\partial f}{\partial \rho}-f$, or the virial equation 
\be
\frac{\beta P}{\rho} = 1 - \frac{\pi\rho}{3}\int_0^{\infty}dr\,r^3 \frac{\partial\beta u(r)}{\partial r} h_{d}(r), 
\ee
where, within the RPA, $h_d^{\rm rpa}(r)=2h^{\rm rpa}(r)$ and $h^{\rm rpa}(r)$ satisfies Eq. (\ref{eq:h_rpa}) for $\lambda=1$.  
The resulting formula is 
\be
\frac{\beta P^{\rm rpa}}{\rho} = 1 + \frac{\varepsilon^*}{2}\bigg[
\frac{{\rm Li}_{5/2}(-\varepsilon^*\eta)-{\rm Li}_{3/2}(-\varepsilon^*\eta)}{\varepsilon^*\eta}\bigg]. 
\ee

\begin{figure}[h] 
 \begin{center}
 \begin{tabular}{rr}
  \includegraphics[height=0.20\textwidth,width=0.24\textwidth]{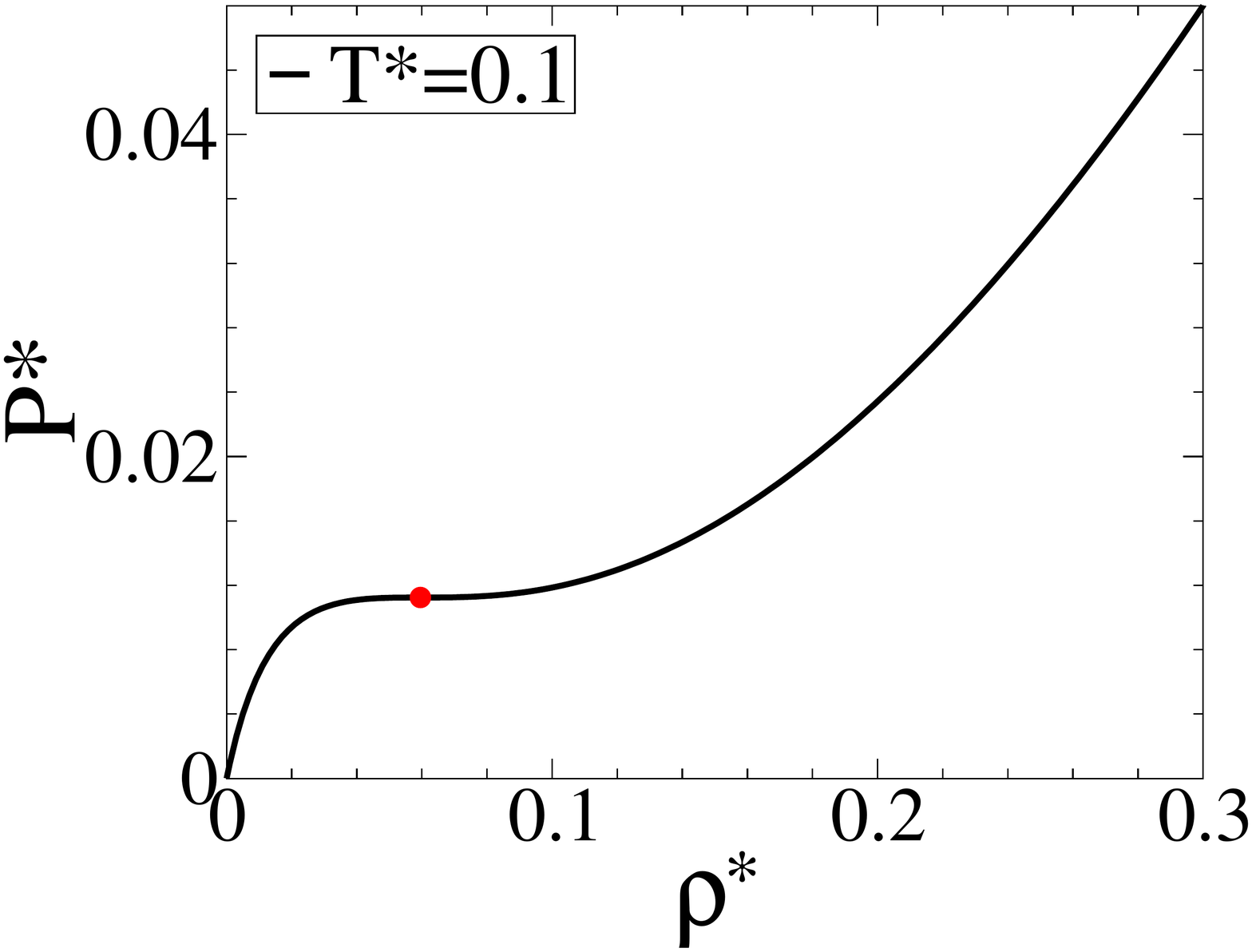}
  \includegraphics[height=0.20\textwidth,width=0.24\textwidth]{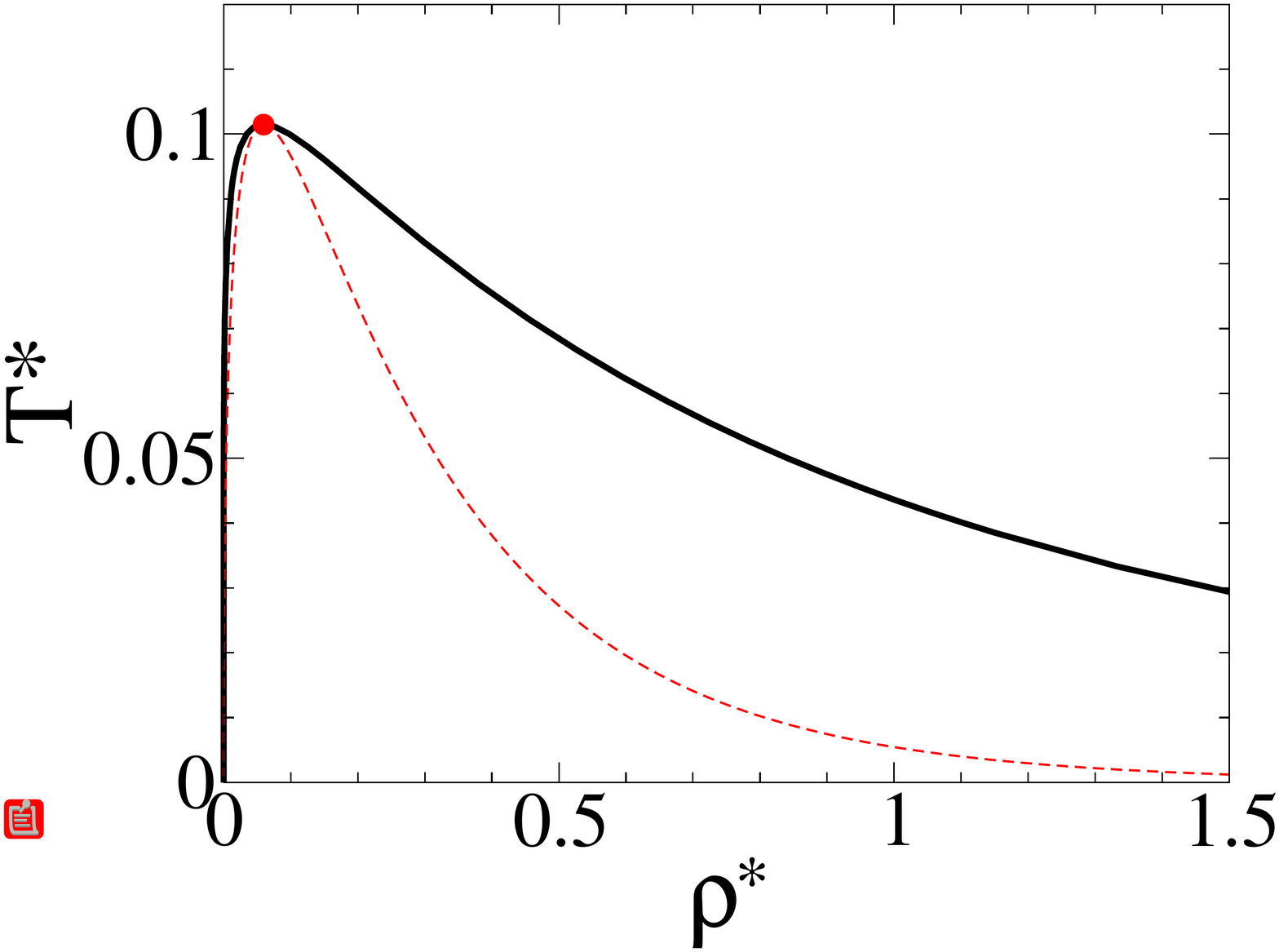}&
 \end{tabular}
 \end{center}
\caption{(a) Pressure isotherm at a critical temperature, $T^*_c\approx 0.1$, from the RPA approximation.  
The red point at $\rho_c^*\approx 0.06$ designates the critical density.  
(b) Phase diagram within the RPA. The spinodal lines 
are indicated with the dashed line, and the coexistence region with the solid line.  } 
\label{fig:P_rpa}
\end{figure}

The critical point is found at $T_c^*\approx 0.1$ (or $\varepsilon_c^*\approx 9.85$) and $\rho_c^*\approx 0.06$,
see Fig. (\ref{fig:P_rpa}) (a).   The entire phase diagram of the 
coexistence region is shown in Fig. (\ref{fig:P_rpa}) (b).  The spinodal lines (designating a metastable region) correspond 
to the local condition $\frac{\partial^2 f}{\partial\rho^2}=0$, while the coexistence region is constructed by the global 
consideration of the free energy, using the Maxwell construction.  
\graphicspath{{figures/}}

The RPA critical temperature is considerably higher than that obtained from simulations, while
the critical density is considerably lower.  To understand some of the causes of this disparity, 
we look into the fluid structure and the behavior of pairs.  
\graphicspath{{figures/}}
\begin{figure}[h] 
 \begin{center}
 \begin{tabular}{rr}
  \includegraphics[height=0.21\textwidth,width=0.27\textwidth]{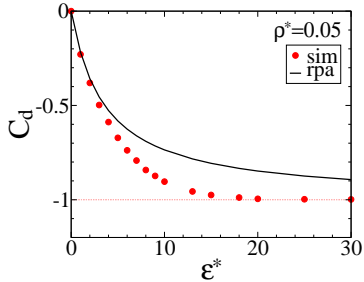}
 \end{tabular}
 \end{center}
\caption{As in Fig. (\ref{fig:C}), but now along with the RPA prediction given in Eq. (\ref{eq:Cd_rpa}). }
\label{fig:C_mf}
\end{figure}
The quantity $C_d$ defined in Eq. (\ref{eq:hc_hole}), within the RPA is 
\be
C_d^{\rm rpa}
 = \rho\hat h^{\rm rpa}(0) 
=-\frac{\varepsilon^* \eta}{1+\varepsilon^* \eta}, 
\label{eq:Cd_rpa}
\ee
where we recall that $\eta=(\sigma\sqrt{\pi})^3\rho$.  
As this algebraic behavior agrees with high temperature results in Fig. (\ref{fig:C}) (b), it fails to agree 
at low temperatures, or the strong-coupling limit, and low density in Fig. (\ref{fig:C}) (a), where an exponential 
decay of $C_d$ to $-1$ indicates the formation of dimers (see Fig. (\ref{fig:C_mf})).  

\begin{figure}[h] 
 \begin{center}
 \begin{tabular}{rr}
  \includegraphics[height=0.20\textwidth,width=0.24\textwidth]{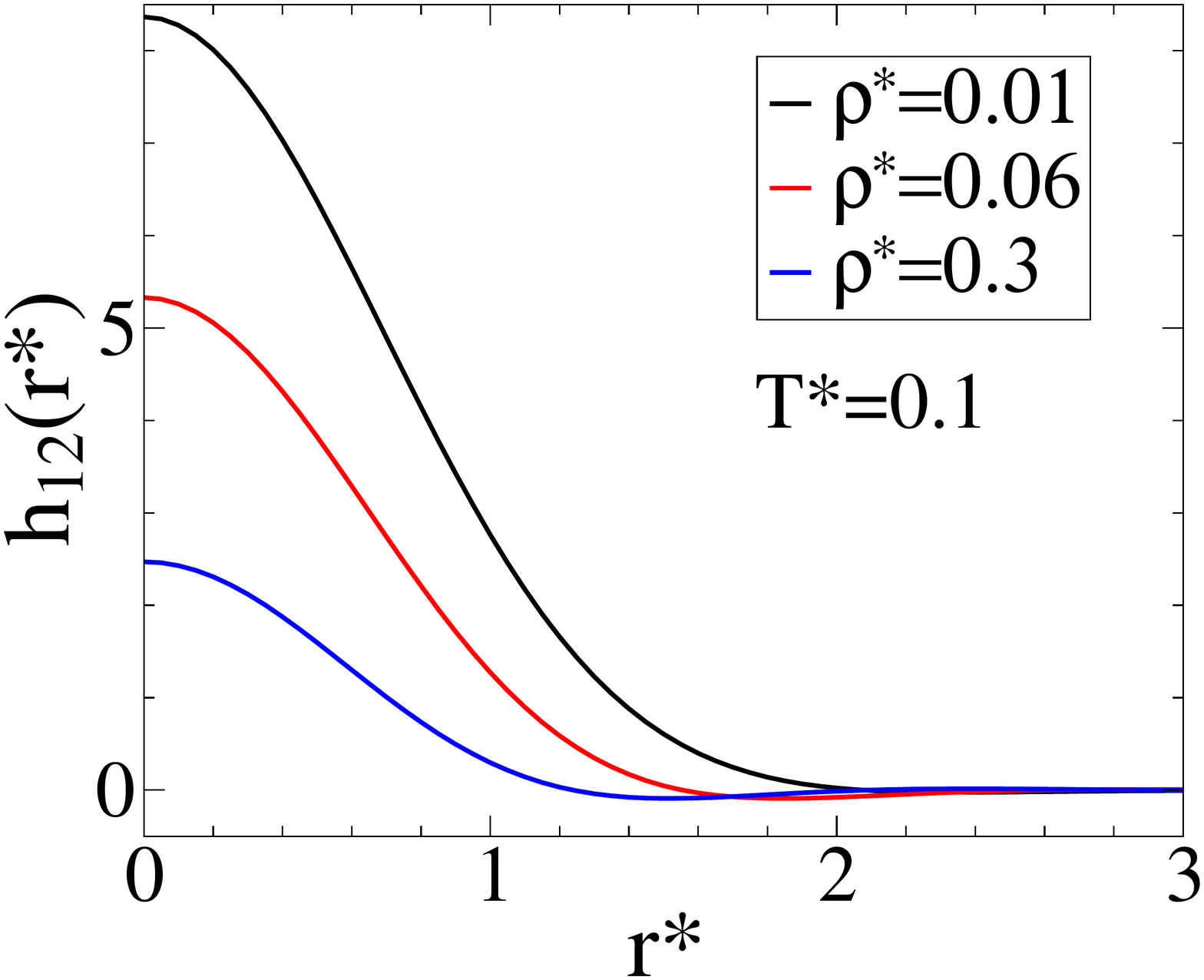}&
  \includegraphics[height=0.20\textwidth,width=0.24\textwidth]{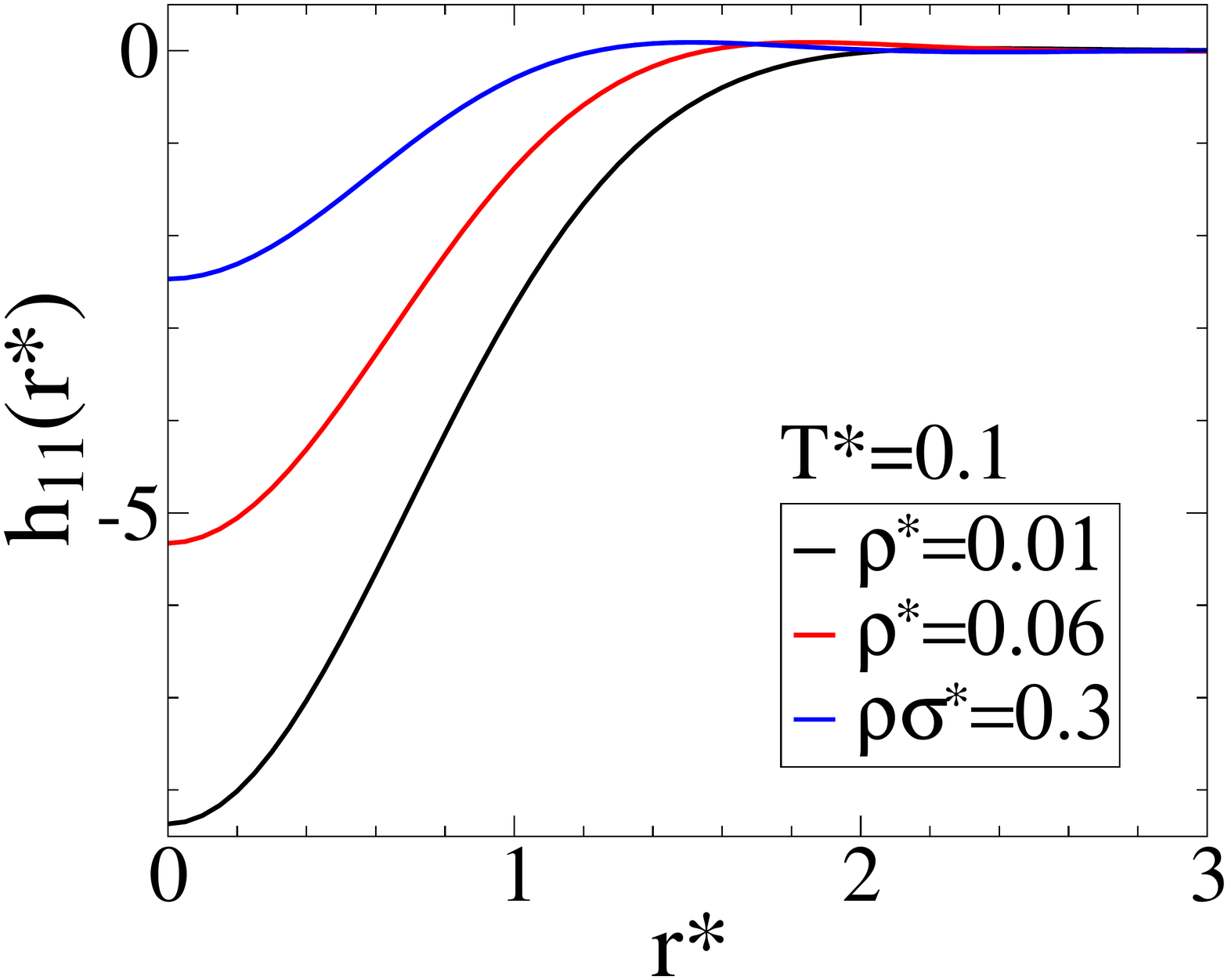}\\
 \end{tabular}
 \end{center}
\caption{Pair correlation functions predicted by the RPA slightly above the critical temperature at $T^*_c\approx 0.1$ for
densities $\rho^*=0.01, 0.06, 0.3$. } 
\label{fig:h_rpa}
\end{figure}

The absence of dimers within the RPA can, furthermore, be attested by from the structure of correlations 
shown in Fig. (\ref{fig:h_rpa}) for $T^*_c\approx 0.1$.  First, the correlations always exhibit oscillatory 
structure.  Second, 
since within the RPA $h_{11}(r)=-h_{12}(r)$, the function $h_{11}(r)$ always exhibits a correlation hole, 
so that the mediated attractive interactions between particles of the same species never arise.
\graphicspath{{figures/}}
This is seen by examining the quantity $C_{11}$, within the RPA given by 
\be
C_{11}^{\rm rpa}
=-\frac{1}{2}\frac{\varepsilon^*\eta}{1+\varepsilon^*\eta},
\ee
which is dominated by negative correlations for any set of parameters.


\subsubsection{RPA in general dimensions}

\begin{figure}[h] 
 \begin{center}
 \begin{tabular}{rrr}
  \includegraphics[height=0.19\textwidth,width=0.24\textwidth]{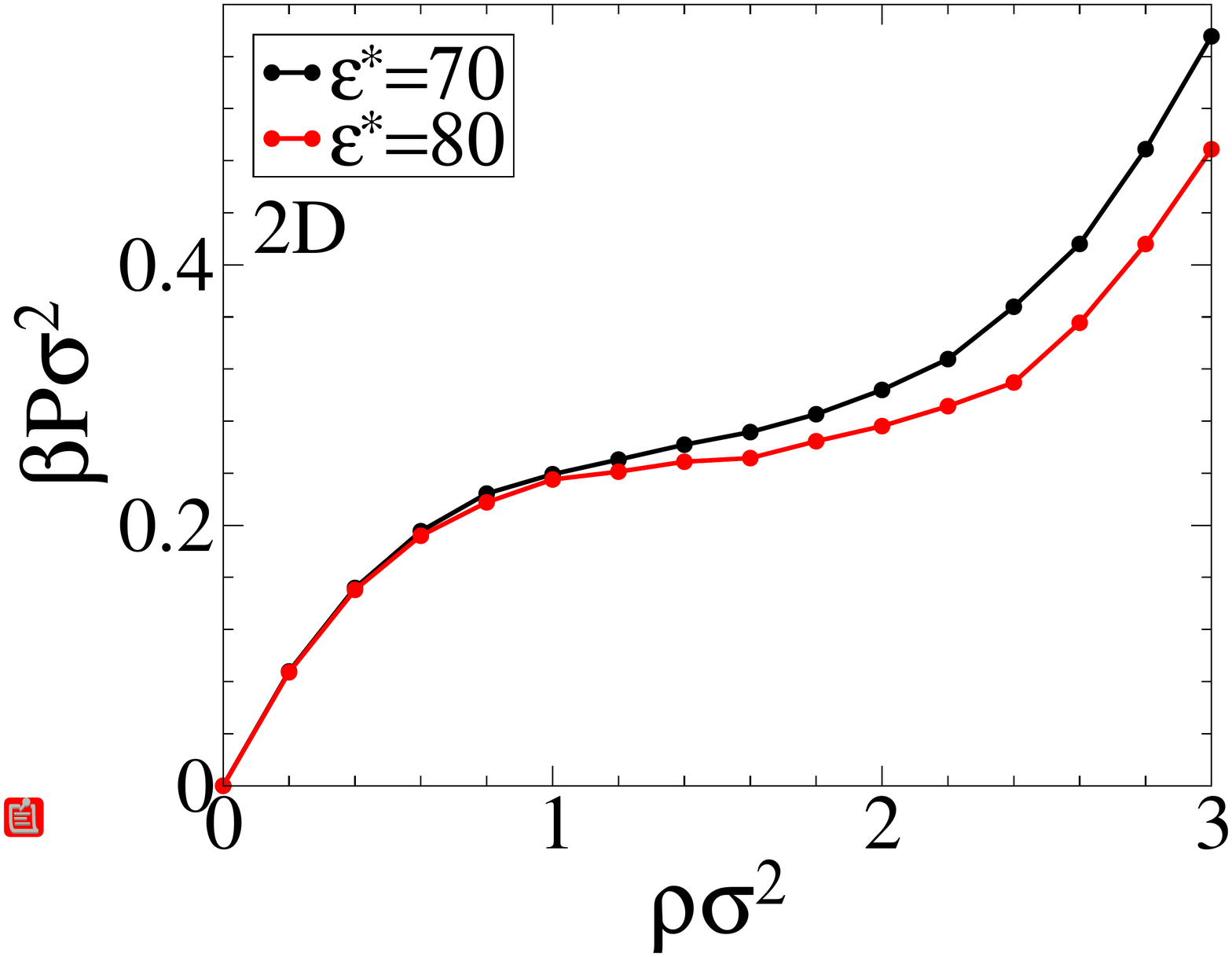}\\
  \includegraphics[height=0.19\textwidth,width=0.24\textwidth]{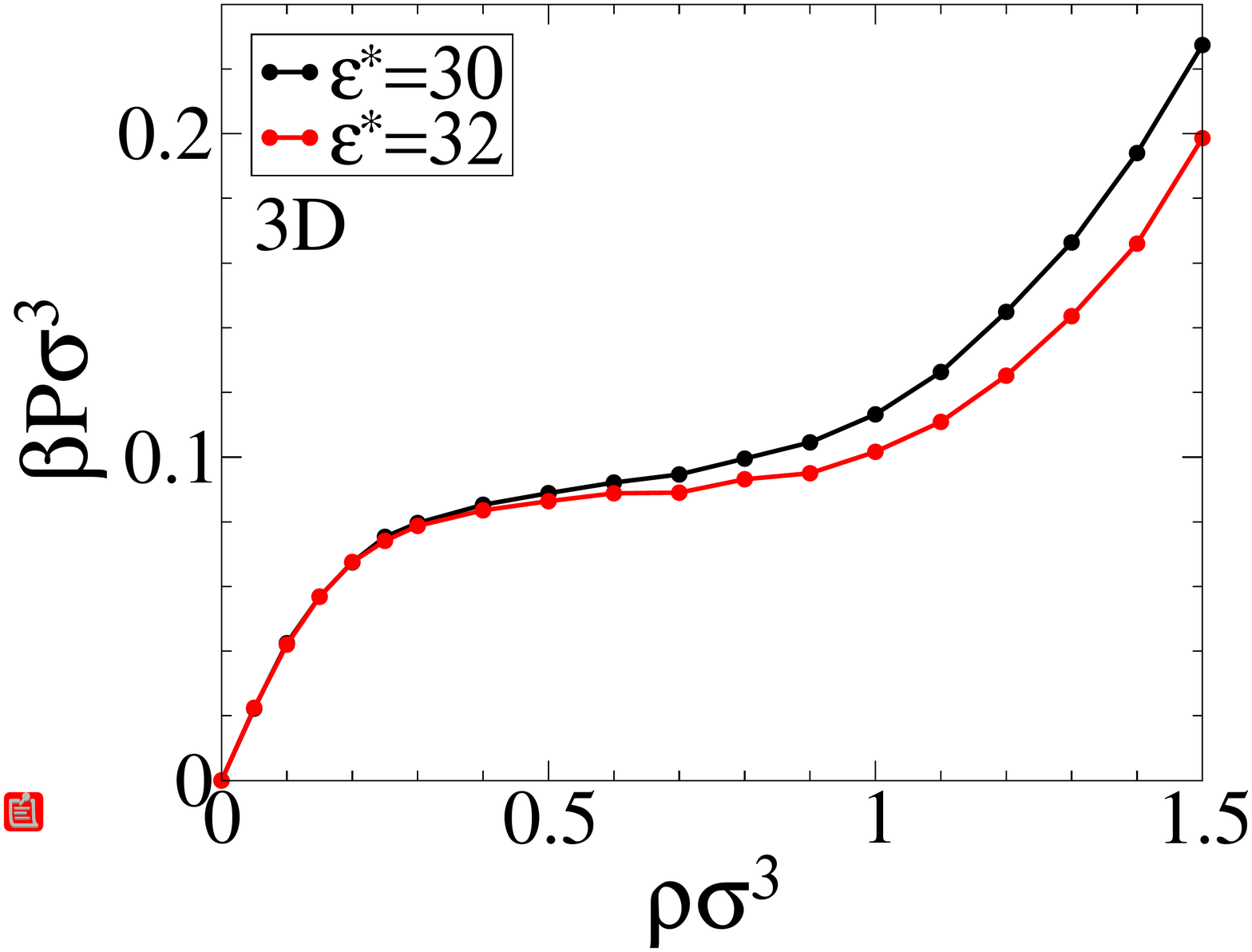}\\
  \includegraphics[height=0.19\textwidth,width=0.24\textwidth]{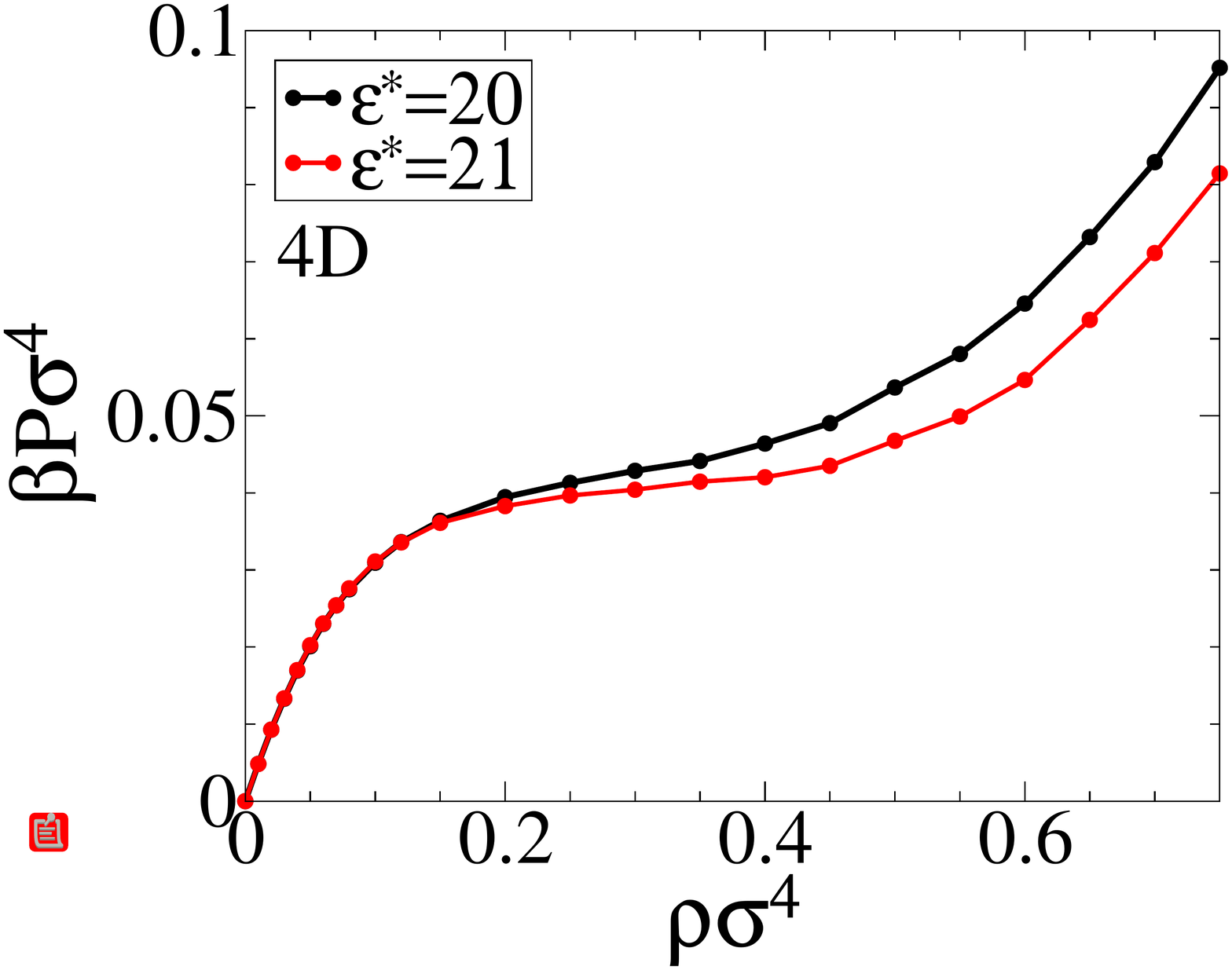}
 \end{tabular}
 \end{center}
\caption{Pressure isotherms for the two-component GCM fluid in various dimensions, $d=2,3,4$, 
slightly above the critical temperature.   
The data points are from the MC simulation for $N=1000$ particles. } 
\label{fig:P_dim}
\end{figure}

In this section we briefly consider the dimension-dependence of the critical point.  
As seen from simulation results in Fig. (\ref{fig:P_dim}), the critical temperature and density decrease 
with reduced dimensionality, as $d=4\to 3\to 2$, and in $d=1$ there is no phase transition.
\graphicspath{{figures/}}

The Fourier transformed Gaussian pair potential for a general dimension is 
$\beta \hat u(k) = (\sigma\sqrt{\pi})^d \varepsilon e^{-k^2\sigma^2/4}$, and the correlational free 
energy becomes 
\ba
\beta f_{\rm c}^{\rm rpa} &=& \frac{1}{(4\pi)^{d/2}\Gamma(d/2)} \nonumber\\
&\times&\int_0^{\infty} dk\, k^{d-1} \bigg(\log\big[1+\rho \beta \hat u(k)\big] - \rho \beta \hat u(k)\bigg), \nonumber
\ea
which evaluates to  
\be
\beta f_{\rm c}^{\rm rpa}  = -\frac{\varepsilon\rho}{2}\bigg\{1+\frac{{\rm Li}_{d/2+1}[-\varepsilon\eta]}{\varepsilon\eta}\bigg\},
\ee
where $\eta = (\sigma\sqrt{\pi})^d\rho$.  Using the thermodynamic definition, 
$\beta P = \rho + \rho\frac{\partial \beta f_{\rm c}}{\partial\rho} - \beta f_{\rm c}$, the pressure is given as 
\be
\beta P = \rho + \frac{\varepsilon\rho}{2}\bigg\{ \frac{{\rm Li}_{d/2+1}(-\varepsilon\eta) - {\rm Li}_{d/2}(-\varepsilon\eta)}{\varepsilon\eta}\bigg\}.  
\ee
\graphicspath{{figures/}}
\begin{figure}[h] 
 \begin{center}
 \begin{tabular}{rr}
  \includegraphics[height=0.22\textwidth,width=0.27\textwidth]{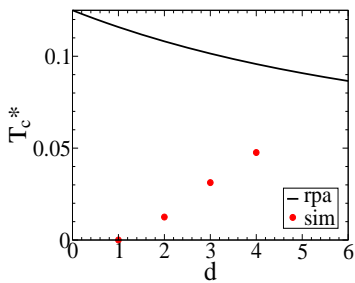}&
 \end{tabular}
 \end{center}
\caption{Critical temperature as a function of dimensionality $d$ for the RPA compared
with the data points from MC simulations.  } 
\label{fig:Tc}
\end{figure}

Critical temperatures from pressure isotherms are 
plotted in Fig. (\ref{fig:Tc})  as a function of $d$, together with the data points obtained from simulations.  
We discover that the RPA fails even in capturing a general trend:  as dimensionality increases, 
the RPA critical temperature decreases.  This shows general inadequacy of the RPA for describing the
strong-coupling limit of two-component systems with interactions of the form presented in Eq. (\ref{eq:u0}).  
In the next sections we consider other possible approaches.  

\subsection{Correlation Functions from the Mean-Field Approximation}
\label{sec:mf}





In the next attempt to treat theoretically the present system, we define a correlation function in terms of a 
density perturbation.  A fixed 
fluid particle of the type "1" at the coordinate origin, generates density perturbations that 
within the mean-field approximation are given by 
\be
\rho_1(r) = \frac{\rho}{2}e^{-\beta u(r)} e^{-\beta w(r)},
\label{eq:rho1_mf}
\ee
\be
\rho_2(r) = \frac{\rho}{2}e^{\beta u(r)} e^{\beta w(r)},
\label{eq:rho2_mf}
\ee
where 
\be
w(r) = \int d{\bf r}'\, u({\bf r},{\bf r}')\big[\rho_1(r') -\rho_2(r')\big], 
\label{eq:w}
\ee
is the mean-potential due to an average distribution of all particles in the system.  
Using the formal definitions  
$\rho_1(r)=\frac{\rho}{2}[h_{11}(r)+1]$ and $\rho_2(r)=\frac{\rho}{2}[h_{12}(r)+1]$, 
the above equations transform into 
\be
h_{11}(r) = e^{-\beta u(r)} e^{-\frac{\rho}{2}\int d{\bf r}'\, \beta u({\bf r}',{\bf r}) [ h_{11}(r')- h_{12}(r')]} - 1,
\label{eq:h11_mf}
\ee
\be
h_{12}(r) = e^{\beta u(r)} e^{\frac{\rho}{2}\int d{\bf r}'\,\beta u({\bf r}',{\bf r}) [h_{11}(r')-h_{12}(r')]} - 1.  
\label{eq:h12_mf}
\ee
Finally, subtracting the two equations,  
we get a single relation 
\be
h_d(r) = -2\sinh\bigg[\beta u(r) +\frac{\rho}{2} \int d{\bf r}'\,\beta u({\bf r}-{\bf r}')h_d(r') \bigg].  
\label{eq:hd_mf2}
\ee
Once the function $h_d(r)$ is obtained from the above self-consistent relation, 
the pressure can be calculated from the virial equation 
\be
\frac{\beta P}{\rho} = 1 - \frac{\pi\rho}{3}\int_0^{\infty}dr\,r^3 \frac{\partial\beta u(r)}{\partial r} h_{d}(r).  
\label{eq:P_vir}
\ee

\begin{figure}[h] 
 \begin{center}
 \begin{tabular}{rr}
  \includegraphics[height=0.21\textwidth,width=0.27\textwidth]{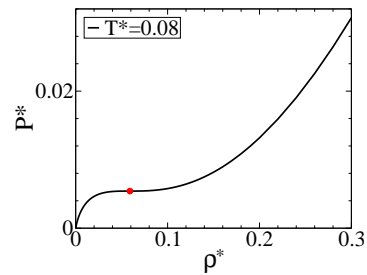}
 \end{tabular}
 \end{center}
\caption{Pressure isotherm at the critical temperature $T^*_c\approx 0.08$, obtained from 
Eq. (\ref{eq:hd_mf2}) and Eq. (\ref{eq:P_vir}). } 
\label{fig:P_mf}
\end{figure}

\begin{figure}[h] 
 \begin{center}
 \begin{tabular}{rrr}
                            \includegraphics[height=0.18\textwidth,width=0.23\textwidth]{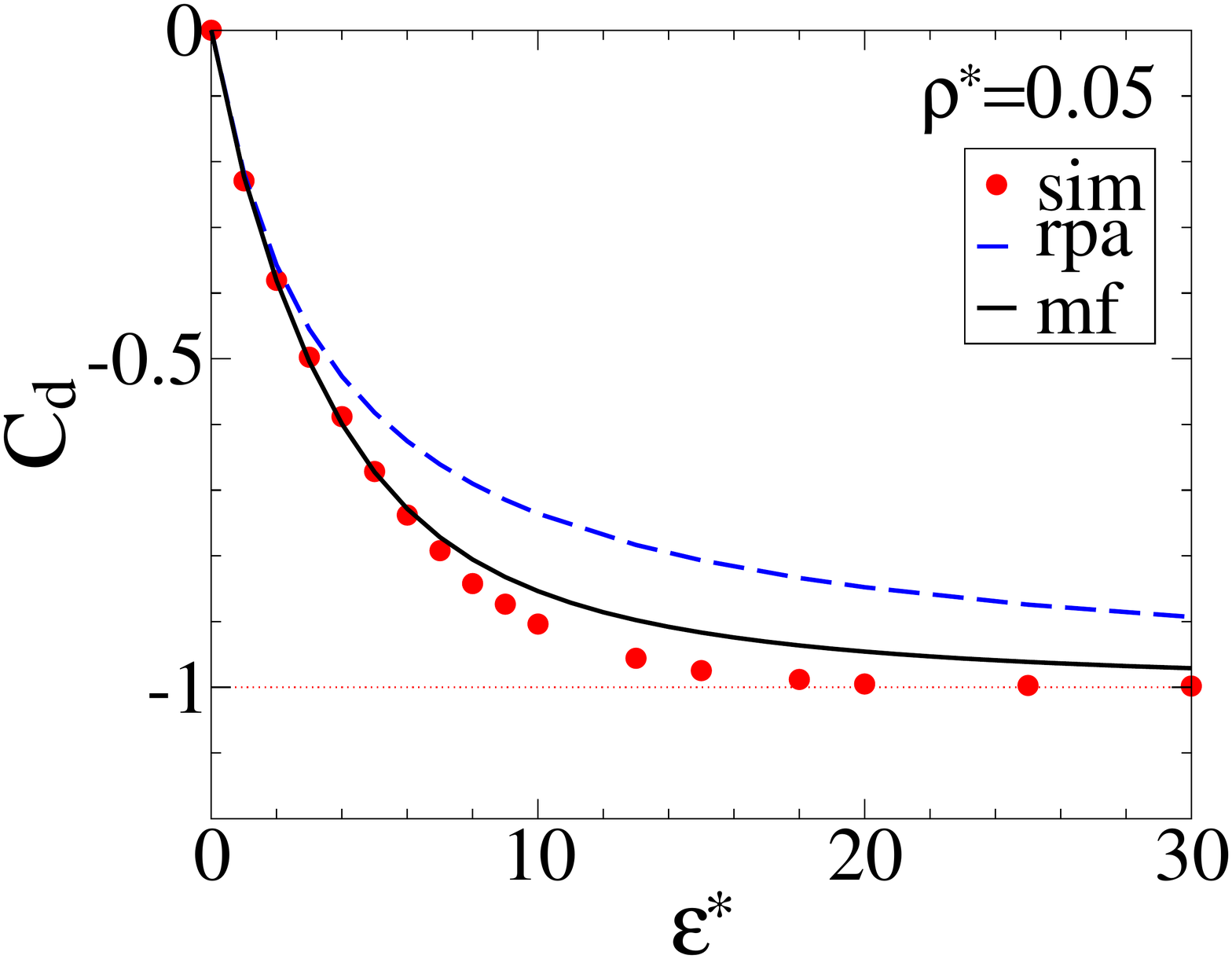}\\
\hspace{-0.2cm}  \includegraphics[height=0.18\textwidth,width=0.23\textwidth]{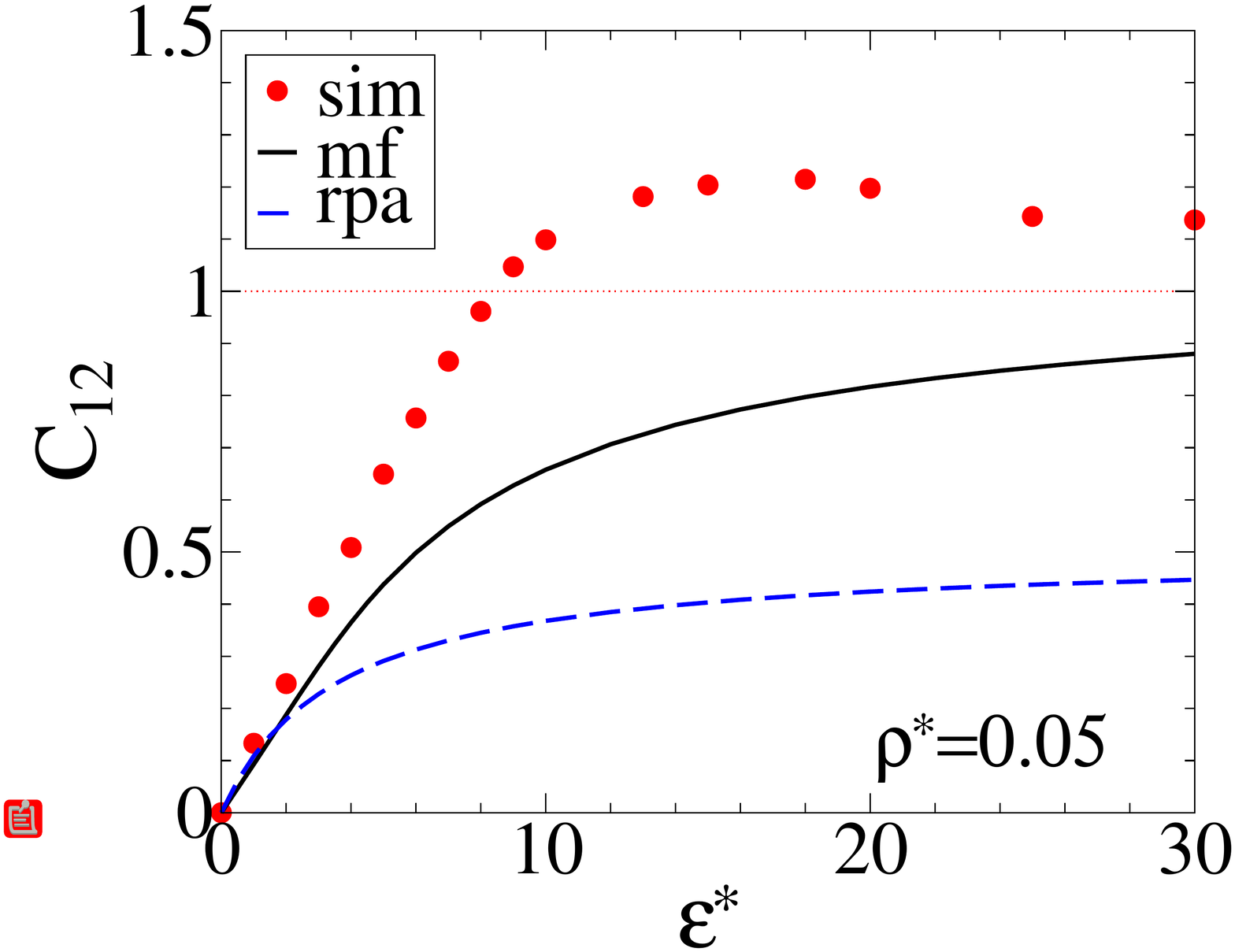}\\
\hspace{-0.2cm}  \includegraphics[height=0.18\textwidth,width=0.23\textwidth]{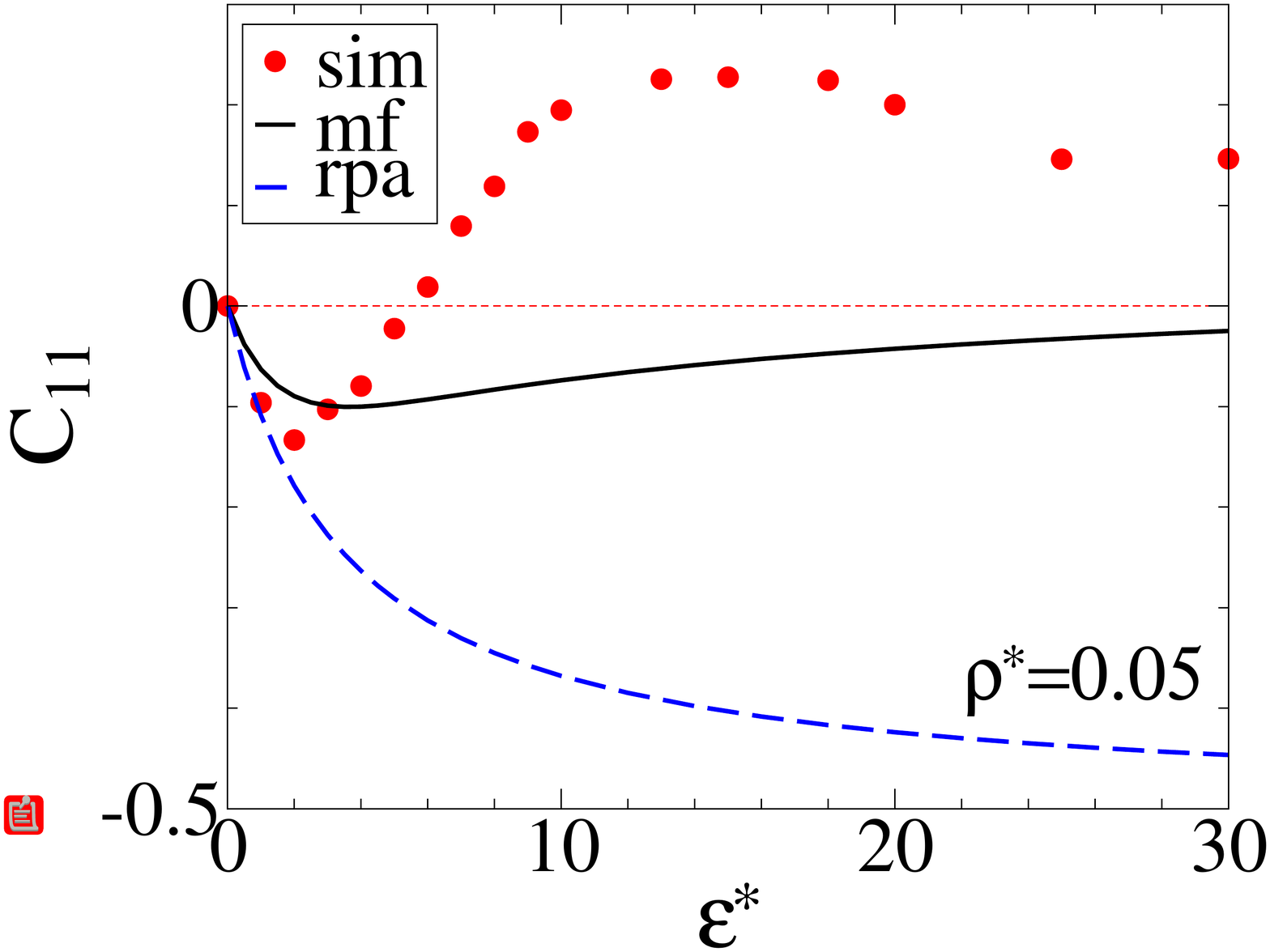}
 \end{tabular}
 \end{center}
\caption{The quantity $C_{ij} = \rho_i \int_0^{\infty}dr\, 4\pi r^2 h_{ij}(r)$ as a function 
of $\varepsilon^*$ for $\rho^*=0.05$. } 
\label{fig:C1_mf}
\end{figure}

In Fig. (\ref{fig:P_mf}) we plot the resulting critical temperature isotherm 
at $T^*_c\approx 0.08$.  
This is slightly lower than that obtained from the RPA approximation ($T^*_c\approx 0.1$ in 
Fig. (\ref{fig:P_rpa})), yet not sufficiently close to the exact result ($T^*_c\approx 0.03$ in 
Fig. (\ref{fig:P_sim})).  The critical density within the present approximation is also slightly shifted,  
from $\rho_c^*\approx 0.06$ within the RPA to $\rho_c^*\approx 0.07$.  
\graphicspath{{figures/}}
In Fig. (\ref{fig:C1_mf}) we plot the quantities $C_{ij}$, in analogy to Fig. (\ref{fig:C}). 
The results show some 
improvement over the RPA, but still there is no indication of pair formation, 
as $C_{11}<0$ and $C_{12}<1$ for all parameters.  
\graphicspath{{figures/}}

\subsection{Correlation Functions from the GRPA Approximation}
\label{sec:grpa}

As in the previous section, we define correlation functions as a perturbation of a uniform fluid 
caused by a fixed particle.  In this section, however, we go beyond the mean-field level of description 
and use the generalized-RPA approximation (GRPA), which is the RPA approximation generalized
to inhomogeneous fluids  \cite{Frydel16}.  


Within the GRPA, the density perturbations caused by a fluid particle of the type "1" fixed at the
coordinate origin are given by 
\be
\rho_1({\bf r}) = \frac{\rho}{2}e^{-\beta u(r)} e^{-\beta w(r)}
e^{\frac{1}{2}[H({\bf r},{\bf r})-H_b(0)]},
\label{eq:rho1_grpa}
\ee
\be
\rho_2({\bf r}) = \frac{\rho}{2}e^{\beta u(r)} e^{\beta w(r)}
e^{\frac{1}{2}[H({\bf r},{\bf r})-H_b(0)]},
\label{eq:rho2_grpa}
\ee
with $w({\bf r})$ defined in Eq. (\ref{eq:w}).  The main difference between these expressions and those 
in Eq. (\ref{eq:rho1_mf}) and Eq. (\ref{eq:rho2_mf}) 
is the presence of a correlation term $H({\bf r},{\bf r})$ obtained from 
\be
H_{}({\bf r},{\bf r}') = -\beta u(r) - \rho \!\! \int \!\! d{\bf r}''\,\big[h_s(r'')+1\big]\beta u({\bf r},{\bf r}'') H_{}({\bf r}'',{\bf r}'),
\ee
which corresponds to the inhomogeneous Ornstein-Zernike equation with the direct correlation function approximated as 
$c({\bf r},{\bf r}')=-\beta u({\bf r},{\bf r}')$, and where $h_s(r)=[h_{11}(r)+h_{12}(r)]/2$.  The quantity $H_b^{\rm }(0)$ 
is the value of $H_{}({\bf r},{\bf r})$ far away from a perturbation, where $\rho_i(r)= \frac{\rho}{2}$.  
Using the formal definitions  
$\rho_1(r)=\frac{\rho}{2}[h_{11}(r)+1]$ and $\rho_2(r)=\frac{\rho}{2}[h_{12}(r)+1]$, the correlation functions become 
\be
h_{11}(r) = e^{-\beta u(r) - \frac{\rho}{2} \int d{\bf r}'\,\beta u({\bf r}-{\bf r}')h_d(r')} e^{\frac{1}{2}[H({\bf r},{\bf r})-H_b(0)]},
\ee
and 
\be
h_{12}(r) = e^{\beta u(r) + \frac{\rho}{2} \int d{\bf r}'\,\beta u({\bf r}-{\bf r}')h_d(r')} e^{\frac{1}{2}[H({\bf r},{\bf r})-H_b(0)]}. 
\ee
Once the correlation functions are calculated, we use Eq. (\ref{eq:P_vir}) to calculate pressure and 
find the critical temperature isotherm.

\begin{figure}[h] 
 \begin{center}
 \begin{tabular}{rr}
  \includegraphics[height=0.21\textwidth,width=0.27\textwidth]{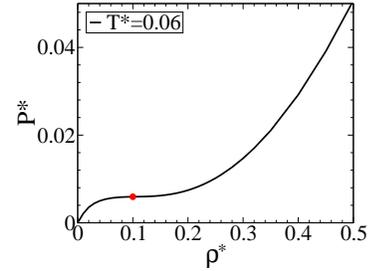}
 \end{tabular}
 \end{center}
\caption{Pressure isotherm at a critical temperature $T_c^*=0.06$ from the GRPA approximation.  } 
\label{fig:P_grpa}
\end{figure}

The predicted critical temperature is $T^*_c\approx 0.06$, and the critical temperature isotherm 
is plotted in Fig. (\ref{fig:P_grpa}), where 
the critical density is $\rho_c^*\approx 0.1$.  This, so far, is the best estimate, at the same time,
it is not accurate enough to be regarded as an accurate theory of the strong-coupling limit.  
\graphicspath{{figures/}}

\begin{figure}[h] 
 \begin{center}
 \begin{tabular}{rr}
\hspace{-0.2cm}  \includegraphics[height=0.19\textwidth,width=0.24\textwidth]{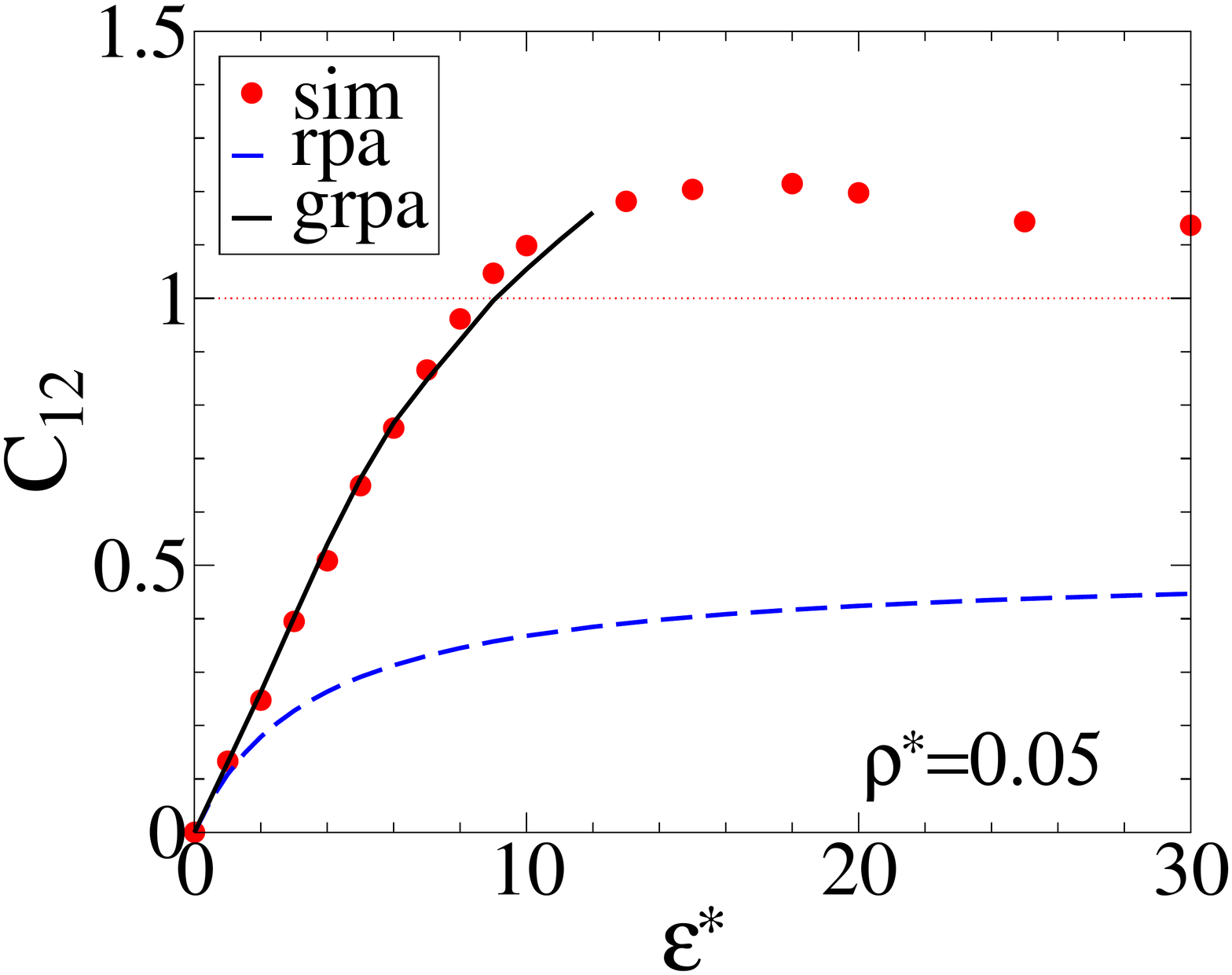}&
\hspace{-0.2cm}  \includegraphics[height=0.19\textwidth,width=0.24\textwidth]{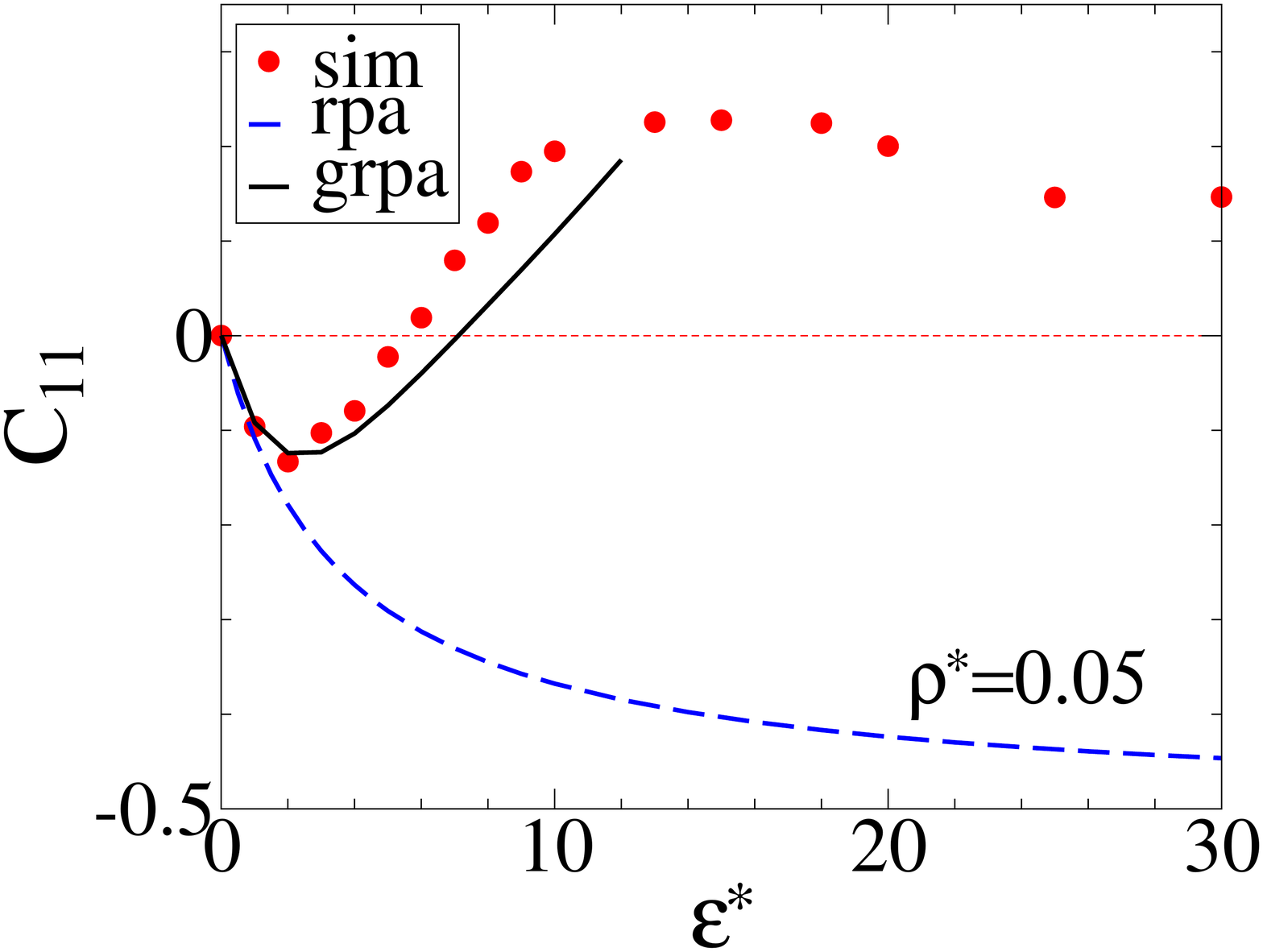}
 \end{tabular}
 \end{center}
\caption{The quantity $C_{ij} = \rho_i \int_0^{\infty}dr\, 4\pi r^2 h_{ij}(r)$ as a function of $\varepsilon^*$ for 
$\rho^*=0.05$.  }
\label{fig:C1_rpa2}
\end{figure}

\begin{figure}[h] 
 \begin{center}
 \begin{tabular}{rr}
  \includegraphics[height=0.19\textwidth,width=0.24\textwidth]{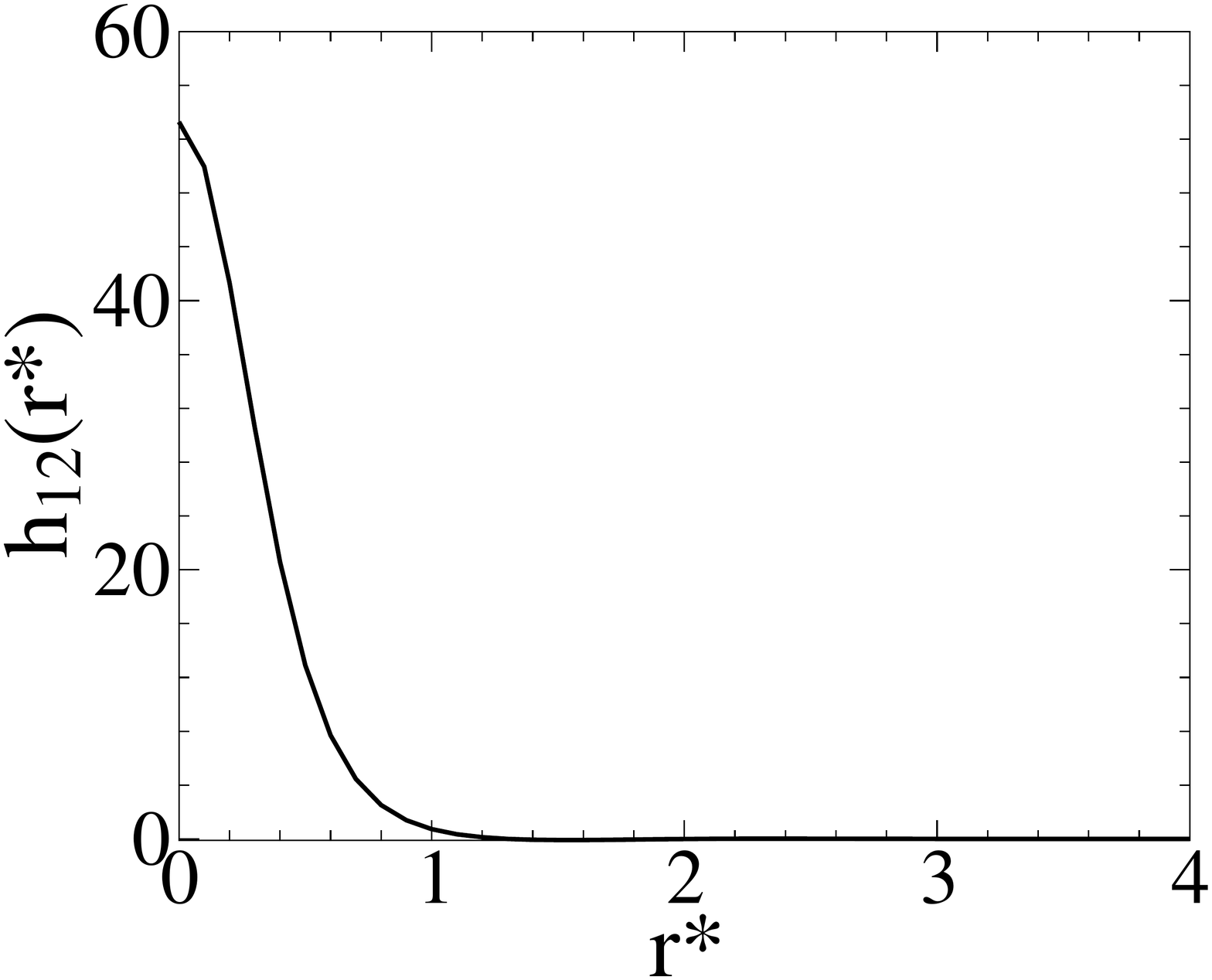}&
  \includegraphics[height=0.19\textwidth,width=0.24\textwidth]{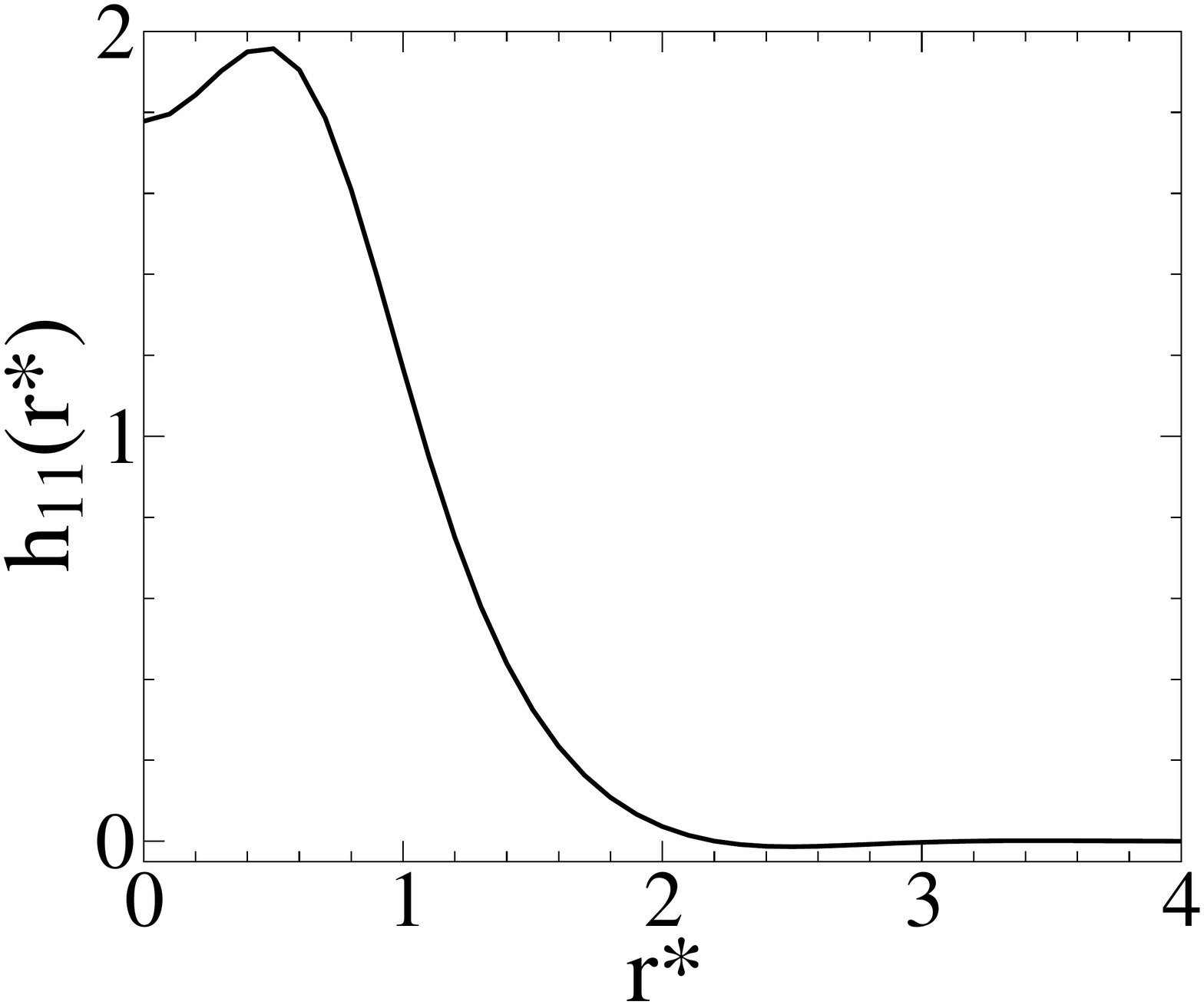}\\
 \end{tabular}
 \end{center}
\caption{Pair correlation functions from the GRPA approximation, 
for $T^*=0.06$ and $\rho^*=0.05$. } 
\label{fig:h_grpa}
\end{figure}

We look next into the quantities 
$C_{ij}$ shown in Fig. (\ref{fig:C1_rpa2}).  Significant feature of the plots is the prediction of 
$C_{11}>0$, and $C_{12}>1$ 
for $\varepsilon^*\gtrsim 8$, indicating mediated attraction between particles of the same
species and, therefore, the presence of dimers. 
\graphicspath{{figures/}}
The existence of dimers if further attested by from the correlation function shown in Fig. (\ref{fig:h_grpa})
for low density and the temperature slightly above $T^*_c$.  
A sharp peak at $r=0$ for $h_{12}(r)$ indicates strong association between particles of the same species,
and absence of the correlation hole and the presence of positive correlations in $h_{11}(r)$ provide additional evidence
for the existence of dimers within the GRPA approximation.  
\graphicspath{{figures/}}

\section{Conclusion}
\label{sec:concl}



A unique feature of the two-component system with the binary interactions 
in Eq. (\ref{eq:u0}) is the special 
role of correlations, which goes beyond a merely correctional role, and provides a  
mechanism for a gas-liquid phase-transition and the formation 
of dimers.  Yet because correlations are dominant only in the strong-coupling limit, 
phase transition and the formation of dimers occur at very low temperatures.  This, in consequence, 
makes theoretical analysis of these phenomena a challenging problem.   

The simplest theory 
of correlations, the RPA, predicts critical temperature at a significantly higher temperature than that
obtained from simulations, and it fails to account for dimer formation.  
The most successful approximation attempted in this work is the GRPA.  This approximation captures 
pair formation, and yields the critical temperature that is closer to the simulation results, yet not 
close enough to be considered an accurate theory.  

In consequence, a theoretical challenge of treating the two-component fluid with interactions in Eq. (\ref{eq:u0})
remains open.  An interesting direction to be considered is to study 
a relevant lattice-gas model, as was done for the one-component GCM fluid in Ref. \cite{Hansen04}.

Finally, based on our results, we do not find evidence that the existence of dimers plays a role in a phase
transition mechanism.   Pairs are prevalent at a low density, let's say $\rho^*<0.1$.  The critical density, on the 
other hand, is at roughly $\rho^*_c\approx0.6$.  At such a high density we no longer find any evidence for the 
existence of dimers, and so the link between pairs and the phase-transition is dubious.  
We regard the formation of dimers and the occurrence of the phase transition 
as two different manifestations of the strong-coupling limit.  To provide support for this 
conjecture, we carried out simulations for permanent dimers (two Gaussian particles of different species 
connected by a spring).  Such a contrived system exhibited no phase transition.  We conclude, therefore, that the
interactions between particles at intermediate and high densities are considerably more complex than that 
provided by the simple reduction to dimers.   

The prediction of dimers, however, can provide a useful test of a performance of an approximation.  
An approximation that predicts dimers can be assumed as appropriate for the strong-coupling limit and
potentially suitable for describing the critical point.

\begin{acknowledgments}
This work was partially supported by the CAPES, CNPq, INCT-FCx, by the US-AFOSR under 
the Grant No. FA9550-16-1-0280, by PNDP-Capes under the project PNPD20132533.  
D.F. would like to acknowledge the usage of computational resources in the ESPCI ParisTech, 
and a kind permission to do so by Tony Maggs and Michael Schindler.  
\end{acknowledgments}



\end{document}